\documentclass[12pt]{article}
\usepackage{graphicx}
\usepackage{bm}
\usepackage{jcapmod}
\usepackage{amsmath}
\usepackage{amssymb}
\usepackage{amsthm}

\newcommand{\be}{\begin{equation}}
\newcommand{\ee}{\end{equation}}
\newcommand{\bea}{\begin{eqnarray}}
\newcommand{\eea}{\end{eqnarray}}
\newcommand{\bdm}{\begin{displaymath}}
\newcommand{\edm}{\end{displaymath}}
\newcommand{\beas}{\begin{eqnarray*}}
\newcommand{\eeas}{\end{eqnarray*}}

\begin{document}

\title{Cosmological Effects of Scalar-Photon Couplings: Dark Energy and Varying-$\alpha$ Models}

\author[\,a]{A. Avgoustidis,}
\author[\,b]{\! C. J. A. P. Martins,}
\author[\,b,c,d]{\! A. M. R. V. L. Monteiro,}
\author[\,b,c]{\! P. E. Vielzeuf,}
\author[\,e]{\! G. Luzzi}

\affiliation[a\,]{School of Physics and Astronomy, University of Nottingham, University Park, Nottingham NG7 2RD, England}
\affiliation[b\,]{Centro de Astrof\'{\i}sica, Universidade do Porto, Rua das Estrelas, 4150-762 Porto, Portugal}
\affiliation[c\,]{Faculdade de Ci\^encias, Universidade do Porto, Rua do Campo Alegre, 4150-007 Porto, Portugal}
\affiliation[d\,]{Department of Applied Physics, Delft University of Technology, P.O. Box 5046, 2600 GA Delft, The Netherlands}
\affiliation[e\,]{Laboratoire de l'Acc\'el\'erateur Lin\'eaire, Universit\'e de Paris-Sud, CNRS/IN2P3, B\^atiment 200, BP 34, 91898 Orsay Cedex, France}

\emailAdd{tavgoust@gmail.com}
\emailAdd{Carlos.Martins@astro.up.pt}
\emailAdd{mmonteiro@fc.up.pt}
\emailAdd{up110370652@alunos.fc.up.pt}
\emailAdd{gluzzi@lal.in2p3.fr}

\date{}
\abstract
{We study cosmological models involving scalar fields coupled to radiation and discuss their effect 
on the redshift evolution of the cosmic microwave background temperature, focusing on links with 
varying fundamental constants and dynamical dark energy.  We quantify how allowing for the coupling 
of scalar fields to photons, and its important effect on luminosity distances, weakens current and future 
constraints on cosmological parameters. In particular, for evolving dark energy models, joint constraints 
on the dark energy equation of state combining BAO radial distance and SN luminosity distance 
determinations, will be strongly dominated by BAO. Thus, to fully exploit future SN data one must also 
independently constrain photon number non-conservation arising from the possible coupling of SN 
photons to the dark energy scalar field. We discuss how observational determinations of the background 
temperature at different redshifts can, in combination with distance measures data, set tight constraints on  
interactions between scalar fields and photons, thus breaking this degeneracy. We also discuss prospects 
for future improvements, particularly in the context of Euclid and the E-ELT and show that Euclid can, even 
on its own, provide useful dark energy constraints while allowing for photon number non-conservation.
}

\maketitle

\section{Introduction} 

The observational evidence for the acceleration of the universe demonstrates that our canonical theories of gravitation and particle physics are incomplete, if not incorrect. A new generation of ground and space-based astronomical facilities (most notably the E-ELT and Euclid) will shortly be able to carry out precision consistency tests of the standard cosmological model and search for evidence of new physics beyond it.

After a quest of several decades, the recent LHC detection of a Higgs-like particle \cite{ATLAS,CMS} finally provides strong evidence in favour of the notion that fundamental scalar fields are part of Nature's building blocks. A pressing follow-up question is whether the associated field has a cosmological role, or indeed if there is another cosmological counterpart. If there is indeed a cosmologically relevant scalar field, the natural expectation is for it to couple to the rest of the degrees of freedom in the model, unless there are symmetry principles suppressing these couplings. Therefore, not allowing for such couplings may significantly bias the analysis of current and future cosmological datasets. 

In this paper, which is a sequel to \cite{Tofz}, we focus on potentially observable signatures of the interaction of cosmological scalar fields with the electromagnetic sector, specifically changes to the standard evolution of the Cosmic Microwave Background temperature with redshift,
\begin{equation}
T(z)=T_0(1+z)\,.
\end{equation}
Rather than focusing on a particular favoured model and obtaining specific constraints, we consider two general classes of 
models from the Cosmology literature, described in a phenomenological way. These include models in which a scalar 
field drives variations of the fine-structure constant, and models where the scalar field is responsible for cosmic acceleration. 
Since scalars can in general couple to the electromagnetic sector, our goal is to demonstrate that such scalar-photon 
couplings, if not accounted for, can strongly bias cosmological parameter determination. We quantify how important an effect 
these couplings can have on constraints derived from current and future datasets, and show how the degeneracy can be broken 
through the independent reconstruction of the CMB temperature evolution.

The two canonical ways to reconstruct the evolution of the CMB temperature with redshift rely on spectroscopy of molecular/ionic transitions triggered by CMB photons \cite{Bahcall,Srianand,Noterdaeme} and on the thermal Sunyaev-Zel'dovich (SZ) effect towards clusters \cite{Fabbri1978,Reph1980,Battistelli,Luzzi}. New sources will soon become available for measurements by the above methods, and entirely new methods for measuring $T(z)$ are also being developed, so it is timely to consider their impact on cosmology and on searches for new physics. There is also complementarity with other techniques that are now becoming available. In \cite{Tofz} we pointed out an important connection with distance-duality tests, allowing to strengthen constrains on models that violate photon number conservation. Such tests have a rich history starting from \cite{Bassett,Uzan}, and a future more precise test will be carried out by Euclid \cite{EuclidTheory}. 

While in \cite{Tofz} deviations from the standard evolution were constrained in a purely phenomenological (but nevertheless model-independent) way, here we discuss in considerably more detail the possible links between $T(z)$ and two other astrophysical observables: measurements of nature's fundamental dimensionless couplings and the equation of state of dark energy. As emphasised in \cite{Pauline1}, such joint measurements will be crucial for the next generation of cosmological experiments, which will carry out precision consistency tests of the underlying scenarios. We will base our discussion on two specific examples, ESA's Euclid\footnote{\url{http://www.euclid-ec.org}} \cite{Euclid} and ESO's European Extremely Large Telescope (E-ELT) \cite{EELT}.  We find that, although for varying-$\alpha$ models determinations of $T(z)$ will not reach the required sensitivity in the near future, in the case of dark energy scalars, their coupling to photons can have a major effect on cosmological parameter determination but the degeneracy can readily be broken with $T(z)$ and distance measurements.

\section{CMB Temperature Evolution} 

We start with a brief review on the redshift evolution of the CMB temperature. This is not meant to be exhaustive, but simply to introduce the basic setup we will work in. Further details can be found in the original analysis of \cite{Lima}, as well as in \cite{Tofz}. We will be assuming the presence of a canonical scalar field in an FRW background, with a $\mathcal{L}$agrangian
\begin{equation}
{\cal L}=\frac{1}{2}(\partial_\mu\phi)(\partial^\mu\phi)-V(\phi)\,,
\end{equation}
with
\begin{equation}
p_\phi=\frac{1}{2}{\dot\phi}^2-V(\phi)\,,\quad \rho_\phi=\frac{1}{2}{\dot\phi}^2+V(\phi)\,.
\end{equation}
Introducing a coupling $C_\phi$ between the scalar field and the radiation fluid, 
the evolution equations for the radiation energy and number densities read
\begin{equation}\label{endens}
{\dot\rho_\gamma}+4H\rho_\gamma=C_\phi\,,
\end{equation}
\begin{equation}\label{numdens}
{\dot n_\gamma}+3Hn_\gamma=\Psi\,,
\end{equation}
where $\Psi$ depends on the coupling $C_\phi$. This will 
in general distort the behaviour of the radiation fluid, and in particular the 
photon temperature-redshift relation, away from its standard evolution. 
Restricting our attention to the observationally relevant case of adiabatic evolution, the adiabaticity condition 
implies~\cite{Lima,Tofz}
\begin{equation}
C_\phi=\frac{p_\gamma+\rho_\gamma}{n_\gamma}\Psi\,,
\end{equation}
and one obtains the following evolution equation for the CMB temperature
\begin{equation}\label{Tdot}
\frac{\dot T}{T}+H=\frac{\Psi}{3n_\gamma}=\frac{C_\phi}{4\rho_\gamma}\,.
\end{equation}
For future use, let us define a correction to the standard behaviour, $y(z)$, such that
\begin{equation}\label{yofz}
T(z)=T_0(1+z)y(z)\,,
\end{equation}
and clearly $y(z)=1$ corresponds to the standard cosmological model; we can then write
\begin{equation}\label{dyovery}
\frac{dy}{y}=\frac{C_\phi}{4\rho_\gamma}dt=-\frac{C_\phi}{4H\rho_\gamma}\frac{dz}{1+z}\,.
\end{equation}

The simplest ansatz for the source term $\Psi$ in Eqs. (\ref{numdens}) and (\ref{Tdot}) is $\Psi=3\beta Hn_\gamma$, 
which yields the relation
\begin{equation}
T(z)=T_0(1+z)^{1-\beta}\,;
\end{equation}
this has been fairly widely used in the past, with the available measurements of $T(z)$ providing a constraint 
on the parameter $\beta$ \cite{Luzzi,Noterdaeme,Tofz}. The corresponding evolution of the radiation density is
\begin{equation}
\rho_\gamma\propto T^4\propto (1+z)^{4(1-\beta)}\propto a^{-4(1-\beta)}\,.
\end{equation}
A generalisation has been suggested by Bassett and Kunz \cite{Bassett}\footnote{In the 
notation of \cite{Bassett}, our $\beta$ corresponds to $\gamma$ and our $\lambda$ 
corresponds to $-\alpha$.}, with $\Psi=3\beta H(1+z)^{\lambda-1}n_\gamma$ and again 
assuming adiabaticity. The previous case is recovered for $\lambda=1$, while for 
$\lambda\neq1$ we get
\begin{equation}
\rho_\gamma\propto a^{-4}\exp{\left[\frac{4\beta}{1-\lambda}a^{1-\lambda}\right]}\,,
\end{equation}
\begin{equation}\label{BK_Tofz}
T(z)=T_0(1+z) \exp{\left[\frac{\beta}{1-\lambda}\left((1+z)^{\lambda-1}-1\right)\right]}\,.
\end{equation}
Naturally, if we linearise in $\beta$ and then in redshift we recover the usual 
linear modification to the standard temperature-redshift relation, $T(z)=T_0[1+(1-\beta)z]$. 
The dependence on $\lambda$ appears at order $\mathcal{O}[\beta(\lambda-2)z^2]$.

For the scalar field energy density we have
\begin{equation}\label{generalscalar}
{\dot\rho_\phi}+3H(1+w_\phi)\rho_\phi=-C_\phi\,,
\end{equation}
which, in models where the scalar field is driving cosmic acceleration, could provide a link 
between temperature evolution (\ref{yofz}-\ref{dyovery}) and the properties of dark energy, 
as will be further discussed below. In particular, for the Bassett-Kunz parametrisation \cite{Bassett} 
for photon loss, mentioned above, $T(z)$ is given by (\ref{BK_Tofz}), while for the dark energy 
density we have:
\begin{equation}\label{darkenergy}
{\dot\rho_\phi}+3H(1+w_\phi)\rho_\phi=-4\beta H a^{1-\lambda}\rho_\gamma\,.
\end{equation}

\section{Links to varying alpha}

In this section we consider models in which an evolving scalar field is driving variations of the fine structure 
constant through its coupling to Maxwell electromagnetism. This also allows photons to be converted into scalar 
particles, effectively violating photon number conservation. The violation can be described by a collision functional 
in the Boltzmann equation, leading to an equation of the form (\ref{numdens}) and modifying the 
temperature-redshift relation through equation (\ref{Tdot}). At the same time, the scalar-photon coupling 
affects luminosity distances in a redshift-dependent way \cite{Tofz}, potentially providing a complementary 
observational channel for probing these models.

For concreteness, let us consider the Bekenstein-Sandvik-Barrow-Magueijo (BSBM) class of models \cite{Sandvik} 
in which the scalar field $\psi$ couples exponentially to the Maxwell $F^2$ term in the matter $\mathcal{L}$agrangian, 
resulting in variations of the fine-structure constant, $\alpha$. In this 
case we have (cf. equation (\ref{endens}))
\begin{equation}
{\dot\rho_\gamma}+4H\rho_\gamma=2{\dot\psi}\rho_\gamma=C_\psi\,,
\end{equation}
with
\begin{equation}
\frac{\alpha}{\alpha_0}=\exp^{2(\psi-\psi_0)}\,,
\end{equation}
and we immediately find, assuming adiabaticity, that (cf. equation (\ref{Tdot}))
\begin{equation}
\frac{\dot T}{T}+H=\frac{1}{2}{\dot\psi}\,,
\end{equation}
\begin{equation}
\ln{\left(\frac{aT}{a_0T_0}\right)}=\frac{1}{2}(\psi-\psi_0)=\frac{1}{4}\ln{\frac{\alpha}{\alpha_0}}\,,
\end{equation}
leading to 
\begin{equation}\label{BSBM_Tofz}
\frac{T(z)}{T_0}=(1+z)\left(\frac{\alpha(z)}{\alpha_0}\right)^{1/4}\sim(1+z)\left(1+\frac{1}{4}\frac{\Delta\alpha}{\alpha}\right)\,,
\end{equation}
which is of the form of equation (\ref{yofz}).
Equation~(\ref{BSBM_Tofz}) was derived for the specific case of exponential coupling appearing in BSBM-type models, 
but note that, since the relative variation $\Delta\alpha/\alpha$ is constrained observationally to be small, such a linear 
dependence of $T(z)$ on $\Delta\alpha/\alpha$ is a good approximation (up to a model-dependent factor of order unity 
allowing for a coefficient different than $1/4$) for a much wider range of couplings. Thus, we can think of (\ref{BSBM_Tofz}) 
as the linear term in a Taylor expansion and use this equation as a more general phenomenological relation that can be tested 
observationally.

Recently, Webb et al. \cite{Dipole} found a significant indication (at the 4.2-$\sigma$ level,) for a spatial dipole in the fine structure  constant, $\alpha$. If this is not a hidden systematic effect, and assuming that this class of models is correct, there should also be an additional CMB temperature dipole (that is, in addition to the standard one) in the same direction of the $\alpha$ dipole, and with $\mu$Kelvin amplitude.  Although this is beyond the scope of the current analysis, it should be possible to disentangle this from the `usual' CMB dipole. In particular, a signal of this magnitude may be of some relevance for the recently released Planck results.

Note that in the above we did not assume an explicit expression for the redshift behaviour of $\alpha$. 
A constant non-zero $\Delta\alpha/\alpha$ would introduce a deviation from the standard temperature-redshift 
law which grows linearly in redshift, see equation (\ref{BSBM_Tofz}). Given that fine-structure constant variations 
are constrained to be weak and that optical/UV spectroscopic measurements lie mostly in the redshift range  
$1\lesssim z\lesssim 3$, taking constant $\Delta\alpha/\alpha$ would be a reasonable approach, and any signal 
for such variations would be a hint of new physics. However, the theory predicts that $\Delta\alpha/\alpha$ should 
evolve and, as observational sensitivity increases (and data at larger redshifts gradually become available), the 
redshift dependence should be included in a model-dependent way. Here, we initiate such an approach, aiming 
at capturing the increase of $\Delta\alpha/\alpha$ with redshift in a phenomenological way. Given the limited 
observational sensitivity we seek a one-parameter phenomenological description for the evolution of 
$\Delta\alpha/\alpha$. However, such a simple parametrisation can be expected to be sufficient because 
most current measurements are firmly in the matter era, where the evolution is simple.    

Let us return to the case of BSBM discussed above. In this class of models, if one neglects the recent 
dark energy domination one can easily find an analytic solution (i.e., a matter era one)
\begin{equation}\label{k_alpha}
\frac{\Delta\alpha}{\alpha}=-4\kappa \ln{(1+z)}\,,
\end{equation}
or in other words
\begin{equation}\label{k_Tofz}
T(z)=T_0(1+z)\left[1-\kappa \ln{(1+z)}\right]\,,
\end{equation}
where $\kappa$ is a dimensionless parameter to be constrained by data. Note that the logarithmic factor 
is of order unity in the observationally relevant redshift range, say $0.5\lesssim z\lesssim 3$.
Dark energy can easily be included numerically but, in this redshift range, it only affects the lower 
end near $z\sim 0.5$ by a factor of order unity. For even smaller redshifts, the effect of dark 
energy is to further damp down variations of $\alpha$ \cite{Sandvik}.  Thus, equation (\ref{k_alpha}) 
is also useful more generally, as a one-parameter toy model alternative to the standard $\beta$ 
parametrisation, which phenomenologically captures the redshift evolution of $\Delta\alpha/\alpha$ 
at intermediate redshifts.

In the following, we will use equation (\ref{k_Tofz}) to make contact with $T(z)$ data at redshifts 
of order unity. We should however first check that values of the parameter $\kappa$ required to produce 
$\alpha$ variations at the level of current sensitivity are not in conflict with 
the atomic clock bounds at $z=0$ \cite{Rosenband}. Since dark energy domination at 
late times has the effect of damping $\Delta\alpha/\alpha$ \cite{Sandvik}, taking
the above redshift dependence for $\alpha$ is conservative in that it overestimates 
$\Delta\alpha/\alpha$ at small redshifts. Assuming (\ref{k_alpha}) we have
\begin{equation}
\frac{1}{\alpha}\frac{d\alpha}{dt}=4\kappa H\,,
\end{equation}
and consequently today we must have
\begin{equation}\label{Deltaalpha0_k}
\left(\frac{1}{\alpha}\frac{d\alpha}{dt}\right)_0 \lesssim 4\kappa H_0=1.3 (\kappa h) \times 10^{-17}s^{-1}\,,
\end{equation}
where the inequality is due to dark energy.
According to Rosenband \textit{et al.} \cite{Rosenband} this variation is constrained to be
\begin{equation}
\left(\frac{1}{\alpha}\frac{d\alpha}{dt}\right)_0=(-1.6\pm2.3)\times 10^{-17} yr^{-1}\,,
\end{equation}
which corresponds to the bound
\begin{equation}\label{kconstraint}
\kappa_{\rm clocks}<(5.4\pm7.7)\times 10^{-8}
\end{equation}
when the inequality in (\ref{Deltaalpha0_k}) is saturated. In practice, $\kappa$ can be significantly larger due to the effect of dark 
energy suppressing $\alpha$ variations in the present era, but note that even this figure is consistent with the aforementioned dipole hint \cite{Dipole}: if one ignores the direction of sources on the sky and naively fits the entire dataset to the above function, there is no strong evidence for a non-zero $\kappa$. The spatial dependence of these measurements will be addressed in subsequent work. There is thus no tension with atomic clock constraints. 

\begin{figure}
\includegraphics[width=8cm]{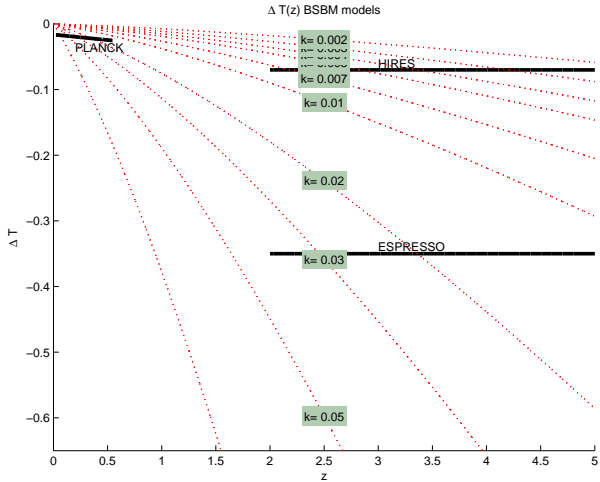}
\caption{Variation of the temperature (relative to the standard model) as function of redshift in a BSBM-like class of models, for different values of $\kappa$ and using $T_0=(2.725\pm0.002)$ K. Also depicted are the limits of detection of this difference with CODEX/HIRES, ESPRESSO and Planck clusters \cite{Tofz}. The span of each bar is meant to represent the typical redshift range of each set of measurements.}
\label{fig1}
\end{figure}

As discussed in \cite{Tofz} the typical precision expected for temperature measurements will significantly increase with the next generation of facilities. Specifically we consider ESPRESSO \cite{ESPRESSO}, under construction for the VLT, and the planned HIRES for the E-ELT (for which the CODEX Phase A study \cite{CODEX} provides a realistic benchmark); their typical expected precisions in the temperature measurements are respectively 
\begin{equation}
\Delta T_{\rm Espresso}\sim 0.35\, K
\end{equation}
and
\begin{equation}
\Delta T_{\rm Hires}\sim 0.07\, K\,.
\end{equation}
These are about three orders of magnitude larger than what one would expect the temperature variation to be in the BSBM model at $z\sim4$, on the assumption that the Webb detection is correct. To get an intuitive picture of the sensitivity of $T(z)$ measurements within this class of models, one can determine what would be the smallest value of $\kappa$ detectable by \textit{a single measurement} by those two future spectrographs. This result is shown in Fig. \ref{fig1}, giving then a detection limit around $\kappa=0.004$ for HIRES and $\kappa=0.02$ for ESPRESSO.

However, these sensitivities will rapidly improve. Note that the detection limits shown in Fig. \ref{fig1} are for a single measurement and they also depend on the redshift at which the measurements are made: the higher the redshift, the stronger the constraints that can be achieved. Clearly a detection of a Webb-level value of $\kappa$ would require a very large number of sources, which are not currently known. However, this may be possible for clusters (whose expected sensitivity in the case of Planck is also depicted): even though they are at much lower redshifts (when deviations from the standard behaviour are correspondingly smaller), samples of thousands of clusters are expected to become available very soon. This is further discussed in \cite{Tofz,DeMartino}.

Recently Muller \textit{et al.} \cite{Muller} have provided the very tight measurement $T=(5.08\pm0.10)$ K at $z=0.89$, using radio-mm molecular absorption measurements. With the ALMA array \cite{ALMA} gradually becoming available, the number and quality of these measurements will steadily increase, although they will be time-consuming and there will be strong competition for ALMA observing time. Nevertheless this method offers the exciting prospect of a new tool to map $T(z)$ over a very wide redshift range, potentially up to $z=6.5$ (J. Black, private communication).

One can also check agreement with the Oklo natural nuclear reactor bound (corresponding to an effective redshift $z=0.14$), which is 
\begin{equation}
\frac{\Delta\alpha}{\alpha}=(0.6\pm6.2)\times 10^{-8}\,
\end{equation}
according to Petrov \textit{et al.} \cite{Petrov}, or
\begin{equation}
\frac{\Delta\alpha}{\alpha}=(0.7\pm1.8)\times 10^{-8}\,
\end{equation}
according to Gould \textit{et al.} \cite{Gould}; this does happen, but in any case there are several caveats with interpreting these Oklo analyses. The most obvious one is that the nuclear reactions being considered are mostly sensitive to the strong nuclear coupling, so assuming that only the fine-structure constant $\alpha$ varies while the rest of the physics is unchanged is a naive assumption. At higher redshifts, an additional consistency test will be provided by the redshift drift measurements carried out by high-resolution spectrographs like HIRES \cite{Pauline1,Pauline2}.

Let us now explore the relation of these measurements with independent determinations of distance measures, in the context of the varying-$\alpha$ models we are considering. In \cite{Tofz} we have shown that if the temperature-redshift relation is changed to
\begin{equation}\label{Tofy}
T(z)=T_0(1+z)y(z)\,
\end{equation}
in models where photon number is not conserved, then the distance duality relation is correspondingly affected:
\begin{equation}\label{distdualviol}
d_L(z)=d_A(z)(1+z)^2y(z)^{3/2}\,,
\end{equation}
where $d_L$ and $d_A$ are the luminosity and angular diameter distance measures respectively. 
Therefore for this class of varying-$\alpha$ models we predict that
\bea
d_L(z)=d_A(z)(1+z)^2 \left(\frac{\alpha(z)}{\alpha_0}\right)^{3/8} &\sim& d_A(z)(1+z)^2 \left(1+\frac{3}{8}\frac{\Delta\alpha}{\alpha}\right)\\
&\sim& d_A(z)(1+z)^2 \left[1-\frac{3}{2}k\ln{(1+z)}\right] \nonumber \,.
\eea
Again, this relation can be tested for both time and/or spatial variations of $\alpha$; even though the effect is small, there are hundreds (and in the future there will be thousands) of type Ia supernova measurements. Recently \cite{Mariano} found a 2$\sigma$ hint for a `supernova dipole' aligned with the $\alpha$ dipole. Their analysis does not take into account the effects of varying $\alpha$ on the supernova brightness (see \cite{Chiba} for a succinct discussion). Such effects are negligible with current supernova sensitivities but they could soon become relevant as datasets of a few thousands of supernovae became available. It would be very interesting to include these effects in a fully consistent analysis.

There are therefore a number of consistency tests for this class of models, involving on the one hand astrophysical measurements of $T$ and $\alpha$, and on the other hand distance measurements such as $d_L$.

\section{Links to dark energy}

We now move to studying models in which a scalar (or pseudo-scalar) field is responsible for dark energy, but also 
couples to the electromagnetic sector. This can happen for example through an axion-like ``F-F dual'' coupling. 
Unlike the case of varying-$\alpha$ models, where we found that current constrains on $\alpha$-variation imply 
that the scalar-photon coupling has a negligible effect on current data interpretation (but could soon 
become significant), in the case of dark energy we will see that, consistent with current constraints, scalar-photon
couplings can have a major impact on cosmological data determinations and must therefore be included in data 
analyses. In particular, the photon-number violation induced by the scalar-photon interactions can significantly 
affect luminosity distances and has an important effect on cosmological analyses that include supernova data. 
However, independent measurements of $T(z)$ can be readily used to break this degeneracy.

Let us consider a cosmological model in which a scalar field $\phi$ is responsible for dark energy, but also 
couples to photons. The coupling involves two photons and a scalar particle and allows, for example, a 
scalar particle to convert into a photon in the presence of a magnetic field. Instead of considering a 
particular model and studying its equations of motion, we will describe this effect in a phenomenological 
way. Our goal is not to study any given model or exhaust all possibilities, but instead to demonstrate that 
such couplings can have a significant effect on cosmology.

The decay of the scalar field can be effectively described in the scalar field fluid equation through a term 
proportional to the scalar energy density. We parametrise this as 
\begin{equation}
{\dot\rho_\phi}+3H(1+w_\phi)\rho_\phi=-3kH\rho_\phi\,,
\end{equation}
where the parameter $k$ (to be constrained) depends on the intergalactic magnetic field. 

\begin{figure}
\includegraphics[width=8cm]{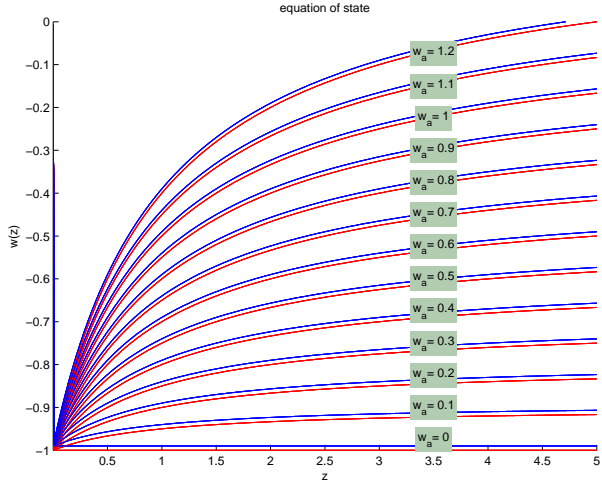}
\caption{The equation of state of dark energy as a function of redshift for different values of the parameters $w_0$ and $w_a$. For each pair of curves the bottom (red) one corresponds to the case $w_0=-1$ and the top (blue) one to $w_0=-0.99$.}
\label{fig2}
\end{figure}

For our present purpose we need to adopt a specific dark energy parametrisation, so we shall take the simple 
and well-known Chevallier-Polarski-Linder (CPL) one \cite{Chevallier}
\begin{equation}
w_\phi(a)=w_0+w_a(1-a)\,.
\end{equation}
Figure \ref{fig2} shows the behaviour of this equation of state for different values of the parameters $w_0$ and $w_a$, chosen so as 
to have an equation of state close to $-1$ at $z=0$ and between $0$ and $-1$ at a redshift of $z=5$. More specifically, we have taken
$w_0=-0.995\pm0.005$ and $w_a=0.6\pm0.6$.  

In this case the scalar field energy density evolves as
\begin{equation}\label{rhogen}
\rho_\phi(a)=\rho_{\phi0}a^{-3(1+w_0+w_a+k)}\exp{[-3w_a(1-a)]}\,;
\end{equation}
in other words the coupling produces an effective correction to the CPL equation of state
\begin{equation}\label{effectivecpl}
w_{\rm eff}(a)=w_{\rm CPL}(a)+k\,.
\end{equation}

It follows that within the context of this class of models we can use $T(z)$ measurements to impose constraints on the dark energy equation of state parameters $w_0$ and $w_a$, as well as on the coupling $k$. This has already been done in two papers by Jetzer \& Tortora \cite{Jetzer,Tortora}, although there it was done for a specific and somewhat unrealistic decaying-lambda model.

If the scalar only couples to radiation (couplings to matter are very strongly constrained), then energy conservation implies that
for the radiation component we have
\begin{equation}\label{rhogamdot}
{\dot\rho_\gamma}+4H\rho_\gamma=3kH\rho_\phi\,.
\end{equation}
Note that this is of the form (\ref{endens}) with $C_\phi=3kH\rho_\phi$.
In this case the evolution equation for the CMB temperature, written in terms of the correction term $y$ defined 
in (\ref{yofz}), is as follows
\begin{equation}
\frac{dy}{y}=\frac{3k}{4}\frac{\rho_\phi}{\rho_\gamma}\frac{da}{a}\,.
\end{equation}
We can then substitute $\rho_\phi$ by the expression above, while for $\rho_\gamma$ we have $\rho_\gamma\propto T^4$. The resulting differential equation for $y$ is not in general analytically integrable, but we can write it as
\begin{equation}\label{ygen}
y^4(a)=1+3k\frac{\Omega_{\phi0}}{\Omega_{\gamma0}}\int_1^a x^{-3(w_0+w_a+k)}\exp{[-3w_a(1-x)]}dx\,,
\end{equation}
which may be integrated numerically. Equation (\ref{ygen}) provides an important observational link between the 
dark energy equation of state and the photon temperature evolution. 

We can obtain analytic approximations in two useful particular limits. First, for a constant equation of 
state (that is, $w_a=0$) and assuming a small $k$ we have
\begin{equation}
y^4(a)\simeq 1+\frac{3k}{1-3w_0}\frac{\Omega_{\phi0}}{\Omega_{\gamma0}} \left[a^{1-3w_0}-1\right]\,,
\end{equation}
which corresponds to
\begin{equation}
T(z)\simeq T_0(1+z)\left[1+\frac{3k}{4(1-3w_0)}\frac{\Omega_{\phi0}}{\Omega_{\gamma0}} \left[(1+z)^{-1+3w_0}-1\right]\right]\,.
\end{equation}

The second (and more specific) limiting case corresponds to small redshifts, $z\ll 1$; here it is convenient to first change variables, integrating in redshift rather than the scale factor, that is
\begin{equation}
y^4(z)=1-3k\frac{\Omega_{\phi0}}{\Omega_{\gamma0}}\int_0^z (1+x)^{3(w_0+w_a+k)}\exp{\left[-3w_a\frac{x}{1+x}\right]}\frac{dx}{(1+x)^2}\,.
\end{equation}
We can now linearise the integral and then integrate, finding
\begin{equation}
y^4(z)\simeq 1-3k\frac{\Omega_{\phi0}}{\Omega_{\gamma0}} z +{\cal O}(z^2)\,,
\end{equation}
which corresponds to
\begin{equation}\label{Tofzsmallz}
T(z)\simeq T_0(1+z)\left(1-\frac{3k}{4}\frac{\Omega_{\phi0}}{\Omega_{\gamma0}} z\right)\,.
\end{equation}
Interestingly, at first order this just depends on $k$, and not on $w_0$ (a $w_0$ dependency does exist at second order).

\begin{figure}
\includegraphics[width=8cm]{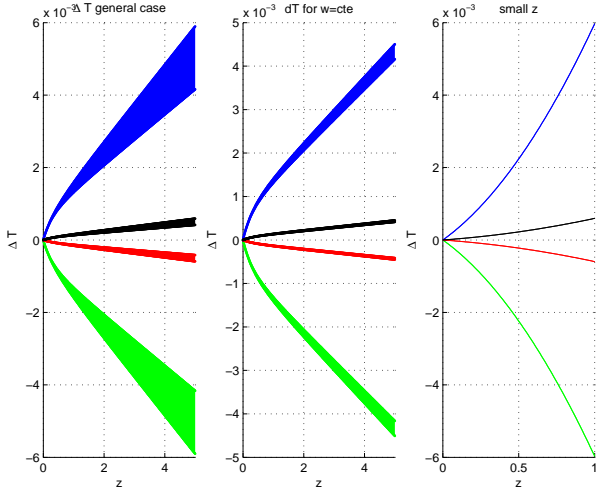}
\caption{Deviation of the temperature relative to the standard model as a function of redshift for the general equation (left), the constant equation of state limit (middle) and the small redshift approximation (right) for different values of the parameters $w_0$,$w_a$ and $k$. From top to bottom each set of curves respectively corresponds to $k=-10^{-7}$, $k=-10^{-8}$, $k=10^{-8}$ and $k=10^{-7}$ respectively; in each case, the corresponding band spans the range of $w_0$ and $w_a$ discussed in the main text. Note that the small-z approximation is only accurate until about $z\sim0.05$.}
\label{fig3}
\end{figure}

Figure \ref{fig3} shows the deviation of the temperature relative to the standard model as a function of redshift for the general case (left), the constant equation of state case (middle) and for the small redshift approximation (right). Given the sensitivities discussed in the previous section, for this parametrisation HIRES will be precise enough to constrain this type of models.

Notice that for sufficiently large values of $k$ (in absolute value) one must have $k$ negative, otherwise the sign of the $y(z)$ factor can change. One can thus infer the prior range of $k$ (as a function of $w_0$) so as to have $y(z)$ non-negative for all redshifts. For the simplest case with $w_a=0$ and with typical values of the other relevant parameters, we find the approximate bound
\begin{equation}\label{kmax}
k<5h^{-2}\times10^{-5}\,.
\end{equation}
In the case of the small redshift approximation (assumed, quite optimistically, to hold up to redshift $z\sim1$) the temperature evolution does not depend on $w_0$, and one immediately finds a limiting value of $k_{max}<10^{-6}$. Another way to set a rough prior range for $k$ is to check whether or not the temperature reaches 3000K in the modified equations. These considerations place $k$ at values too small to make a significant contribution on the effective equation of state of dark energy in equation (\ref{effectivecpl}). However, the effect of $k$ on the temperature evolution, equation (\ref{ygen}), can be large as it is enhanced by the factor $\Omega_{\phi0}/\Omega_{\gamma0}$. As we will see in the next section, the corresponding effect on luminosity distances for values of $k$ at this level can be very large.

As before, it is useful to get an intuitive idea for the sensitivity of $T(z)$ measurements on the dark energy equation of state in this class of scalar field models. Assuming that $w_a=0$ and the present matter density of the universe is known, we find that a set of ESPRESSO $T(z)$ measurements could on its own constrain $w_0$ to a precision $\delta w_0\sim0.4$, while a HIRES-like spectrograph can reach $\delta w_0\sim0.15$; these numbers apply for an optimistic choice of fiducial model with $k\sim-10^{-5}$. (Note that these constraints are weaker than those found by \cite{Jetzer,Tortora}, but that's due to the fact that these authors are assuming a specific model were $w_0$ and $k$ are not independent, i.e. they have a single free parameter to constrain.) On the other hand, the analysis does not include $T(z)$ measurements from clusters or from ALMA, which will be discussed in more detail elsewhere.

As a final comment we point out that the same methods can be applied to constrain a wide range of phenomenological models
of the form (\ref{endens}),(\ref{generalscalar}). As an example, take
\begin{equation}
{\dot\rho_\phi}+3H(1+w_\phi)\rho_\phi=-4\beta H\rho_\gamma\,,
\end{equation}
which is the particular case of Bassett and Kunz \cite{Bassett} for $\lambda=1$, cf. Eq. (\ref{darkenergy}). In this case the 
evolution of the radiation density and its temperature are trivially
\begin{equation}
T(z)=T_0(1+z)^{1-\beta}\,
\end{equation}
\begin{equation}
\rho_\gamma\propto T^4\propto (1+z)^{4(1-\beta)}\propto a^{-4(1-\beta)}\,,
\end{equation}
as before, but we have more complex evolution for the dark energy density: the dark energy equation of state effectively gets a $\beta$-correction, and therefore a constraint on $\beta$ may be inferred, for example, from combining type Ia supernova measurements with other  distance measure determinations probing the cosmic expansion history.

\section{Constraints from current data\label{current}}

Before discussing in more detail some prospects for the next generation of relevant observational 
facilities, we study the constraints that can be obtained from current data. These are already useful, even 
for the general case given by equations (\ref{rhogen}), (\ref{ygen}) and (\ref{Tofy}).

The evolution of the dark energy density (\ref{rhogen}) affects cosmic 
expansion (predominantly at smaller redshifts $z\lesssim 1$ when dark energy starts to dominate) 
so all distance measures depend explicitly\footnote{Recall that the dependence on $k$ is negligible, 
as this is allowed to be at most $\sim10^{-5}$, while $w_0$ and $w_a$ range over intervals of order unity.} 
on $w_0$ and $w_a$.  We can use, for example, type Ia Supernova measurements (giving $d_L(z)$), 
BAO (yielding $H(z)$ and $d_A(z)$), galaxy ageing (providing independent measurements of $H(z)$) 
and $H_0$ determinations. On the other hand, equations (\ref{ygen}) and (\ref{Tofy}) have a strong 
dependence on $k$ and a different dependence on $w_0$, $w_a$, allowing degeneracies to be broken. 

As was pointed out above, in the models we are considering, in which the deviation from the standard 
$T(z)$ relation is due to a coupling of CMB photons with the dark energy scalar field, one generally expects 
that the same field also couples to optical photons, thus affecting luminosity distances, as discussed in \cite{Tofz}.
Within a given model, one can then translate $T(z)$ 
deviations to violations of the distance duality relation (\ref{distdualviol}). Note that on general grounds, 
the coupling is expected to be weaker for lower photon frequencies, so assuming a frequency-independent 
coupling should yield conservative bounds on $T(z)$ violations from SN (or other optical) data. 

We use the Union2.1 SNIa compilation \cite{Union2p1} and a number of different determinations of $H(z)$: 
cosmic chronometers \cite{Stern2009,Simon2004,Jimenez2003} (11 data points in the redshift range $0.1<z<1.75$) 
and the more recent \cite{Moresco2012} (8 data points at $0.17<z<1.1$), 
BAO combined with Alcock-Paczynski (AP) distortions to separate the radial component 
in the WiggleZ Dark Energy Survey \cite{Blake2012} (3 data points at $z=0.44,0.6$ and $0.73$), the 
SDSS DR7 BAO measurement \cite{Xu2012} at $z=0.35$, the BOSS BAO+AP measurement 
\cite{Reid2012} at $z=0.57$, 
and the recent $H(z)$ determination at z=2.3 from BAO in the Ly$\alpha$ 
forest of BOSS quasars \cite{Busca}. This gives 25 data points in the range $0.1<z<2.3$.    

\begin{figure*}
\includegraphics[width=8cm]{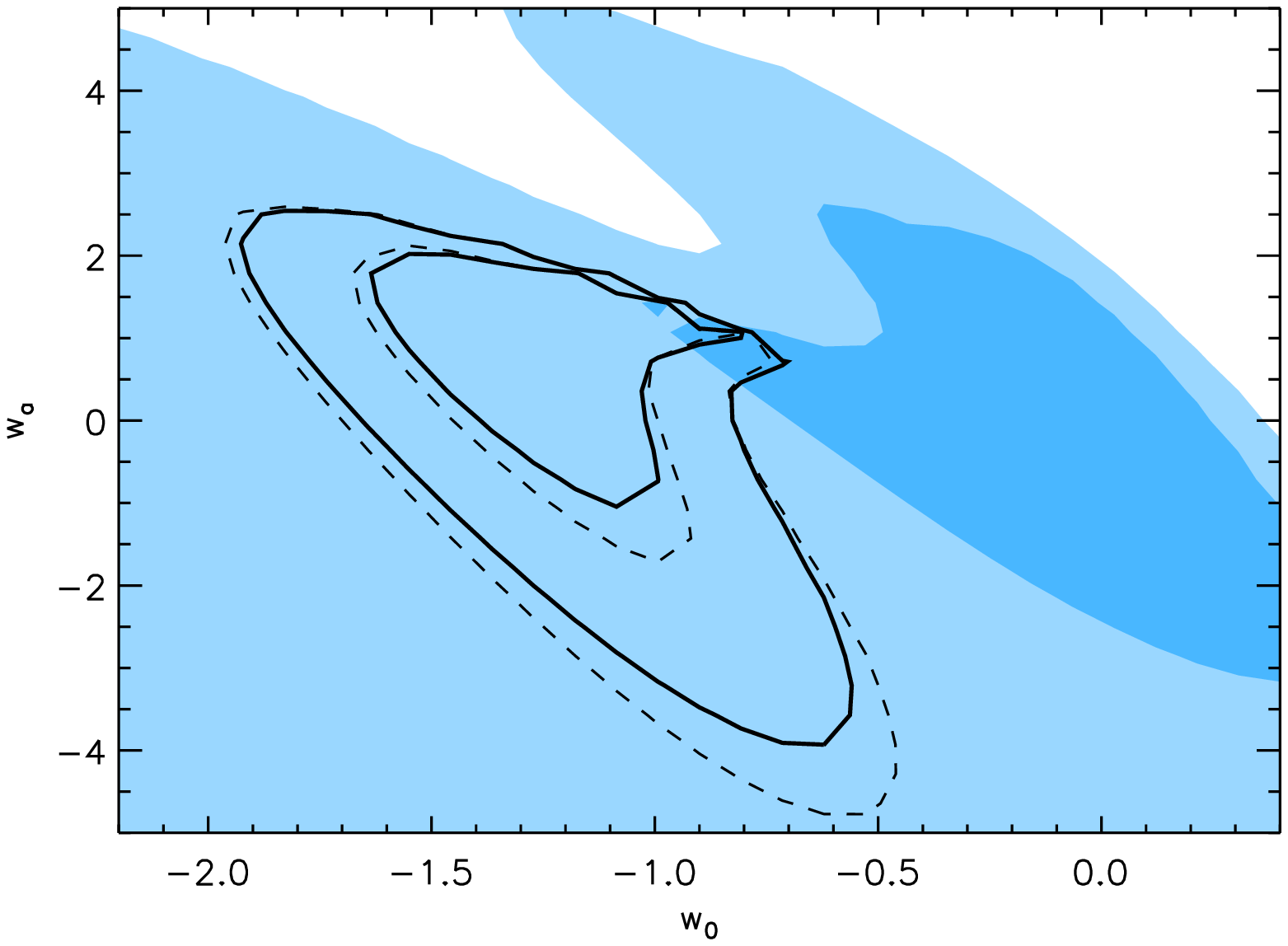} 
\includegraphics[width=8cm]{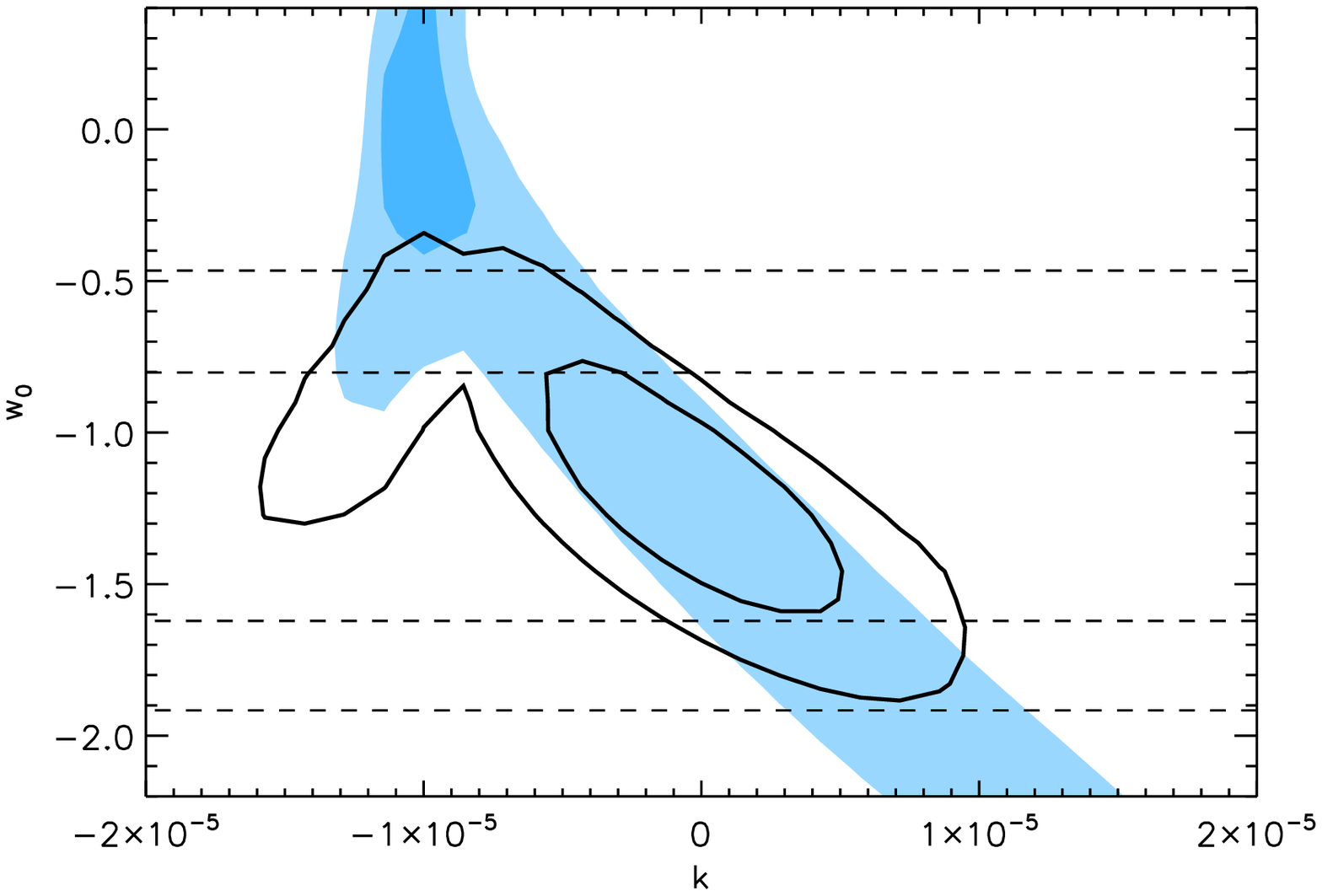}
\includegraphics[width=8cm]{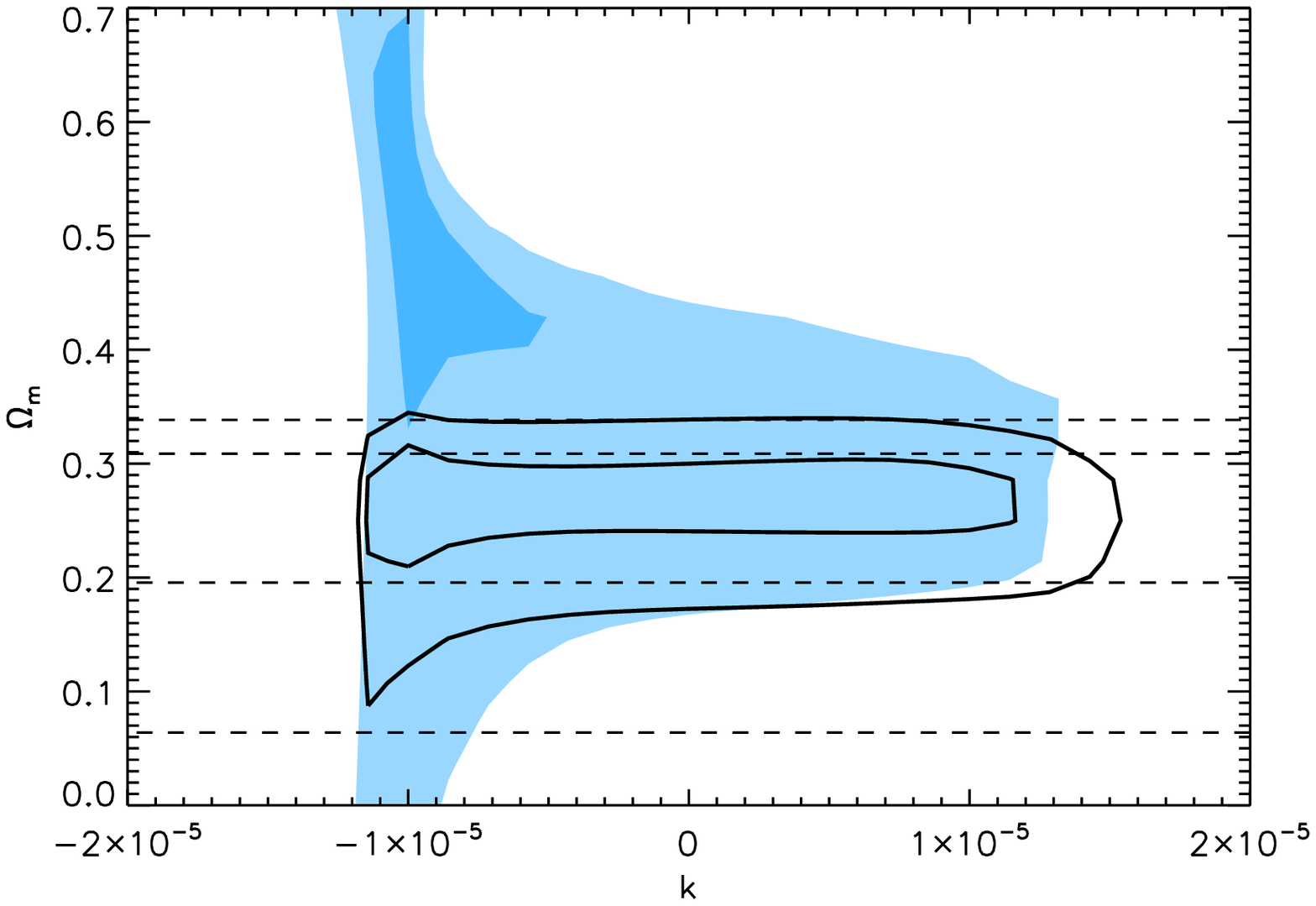}  
\includegraphics[width=8cm]{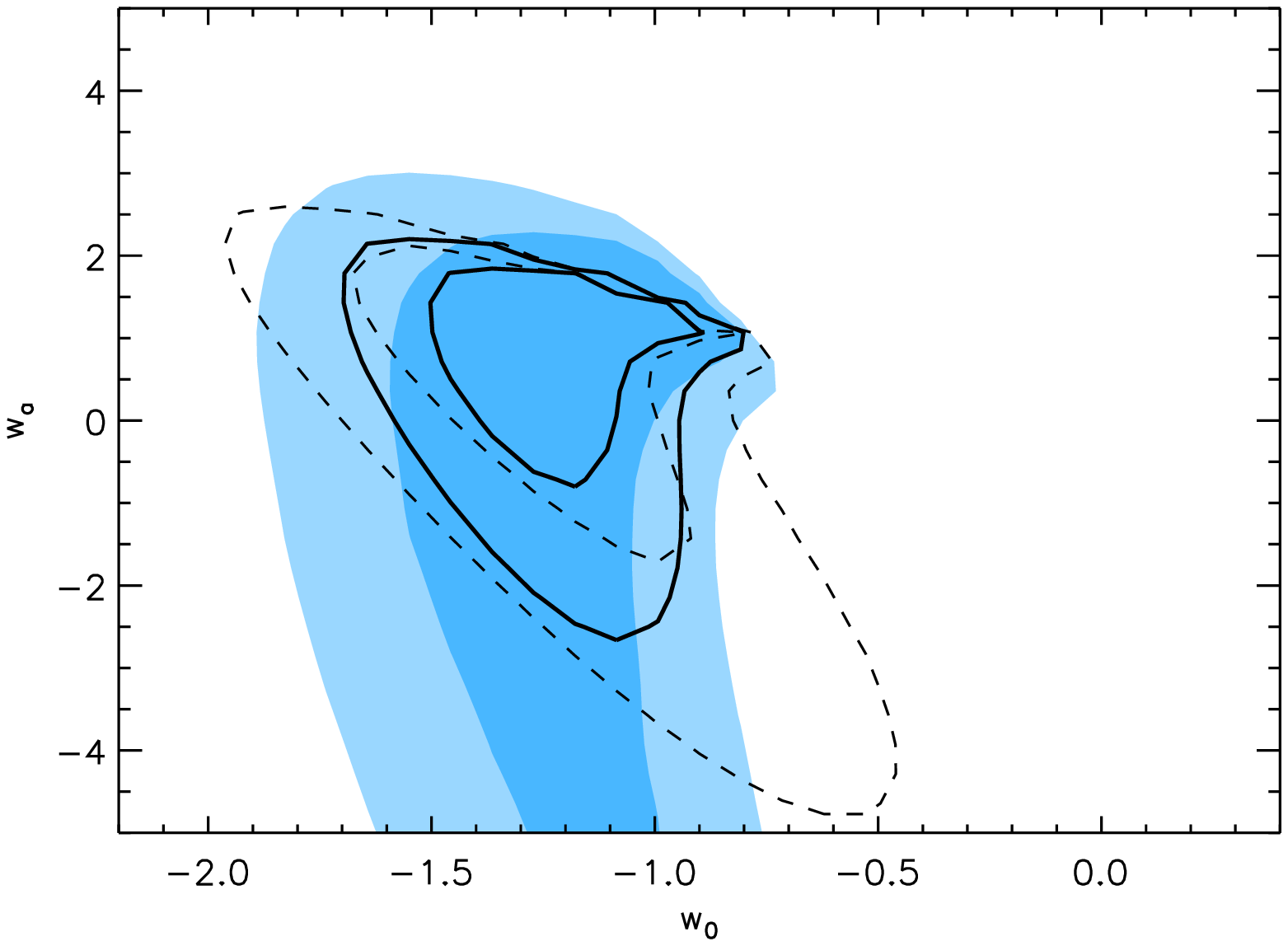} 
\caption{Current constraints on flat CPL-CDM models, allowing 
for violation of photon number conservation, parameterised by $k$. 
\emph{Top Left:} 2-parameter joint constraints (68\% and 95\%, having 
marginalised over $k$ and $\Omega_m$) for the SN (blue filled contours), 
H(z) (dashed lines) and combined SN+H(z) data (solid line contours). Having 
allowed for photon number non-conservation, the SN data alone do not strongly
constrain dark energy (and favour $w_0\sim 0$ at the $1\sigma$ level), but 
the inclusion of the H(z) data strongly improves dark energy constraints.
\emph{Top Right-Bottom Left:} The SN data favoured region $w_0\sim 0$ 
corresponds to negative $k$ (photon dimming) and large $\Omega_m$, 
exemplifying the well-known dark energy-photon dimming degeneracy. This 
gets broken by using the H(z) data which favour $\Omega_m\simeq 0.25$.    
\emph{Bottom Right:} As in top left plot but now assuming that photon-number 
is conserved $k=0$. Note the dramatic effect of $k$ on SN constraints 
(blue filled contours).}\label{conservpriors}
\end{figure*}

We start with a conservative choice of (flat) priors, namely $\Omega_m\in [0,1]$, $w_0\in[-2.2,0.4]$, 
$w_a\in[-5,5]$ and $k\in[-5,5]\times 10^{-5}$. For our $H(z)$ analysis, we marginalise over $H_0$ assuming 
a Gaussian prior based on the Riess et al \cite{Riess2011}  determination $H_0=73.8 \pm 2.4$ km/s/Mpc, while 
in our SN likelihoods we effectively marginalise over $H_0$ by marginalising over intrinsic SN brightnesses. 
Fig. \ref{conservpriors} shows current constraints from combining the above luminosity distance 
and radial distance data on our (flat) CPL-CDM models, allowing for a non-zero $k$ in equation (\ref{rhogamdot}).
The top left panel of Fig. \ref{conservpriors} shows 2-parameter joint constraints (68\% and 95\%, having marginalised 
over $k$ and $\Omega_m$) for the SN (blue filled contours), H(z) (dashed lines) and combined SN+H(z) 
data (solid line contours). Having allowed for violation of photon number conservation through the parameter 
$k$, the SN constraints on the dark energy equation of state are weak, but the constraints improve dramatically 
with the inclusion of $H(z)$ data that are not affected by $k$. The region near $w_0=0$, favoured (at the 1$\sigma$
level) by the SN data, corresponds to negative values of $k$ (i.e. photon dimming due to decay to scalar particles) 
as is evident from the top right panel of Fig. \ref{conservpriors}. The bottom left panel then shows that corresponds 
to large values of $\Omega_m$, so low values for the dark energy density parameter. This is the well-known 
degeneracy between dark energy and photon number non-conservation, which gets broken by including the 
$H(z)$ data favouring $\Omega_m$ near 0.25. For comparison, in the bottom right panel we show the 
corresponding constraint on the $w_0-w_a$ plane but now assuming that photon-number is conserved, $k=0$.       

\begin{figure*}
\includegraphics[width=8cm]{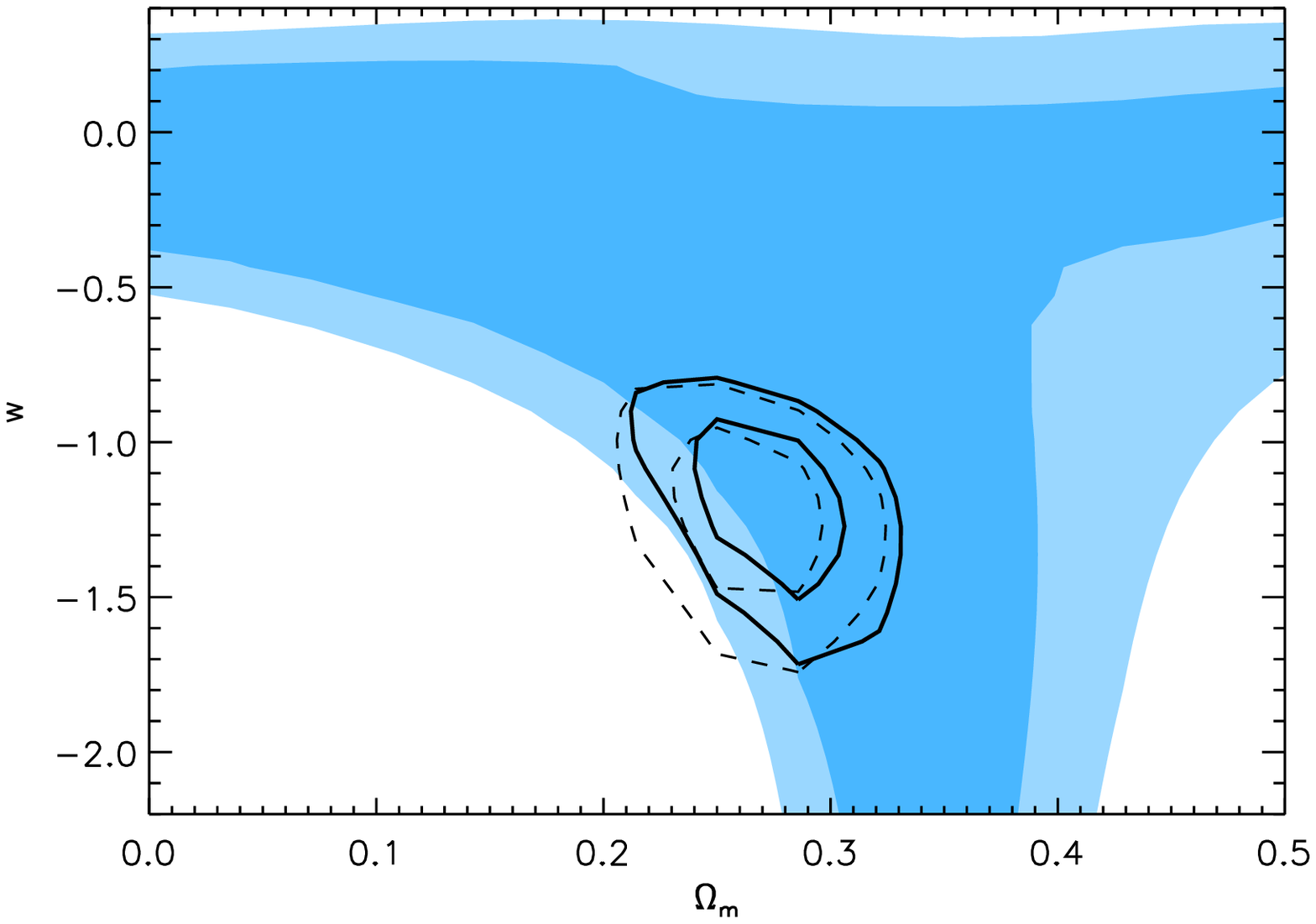} 
\includegraphics[width=8cm]{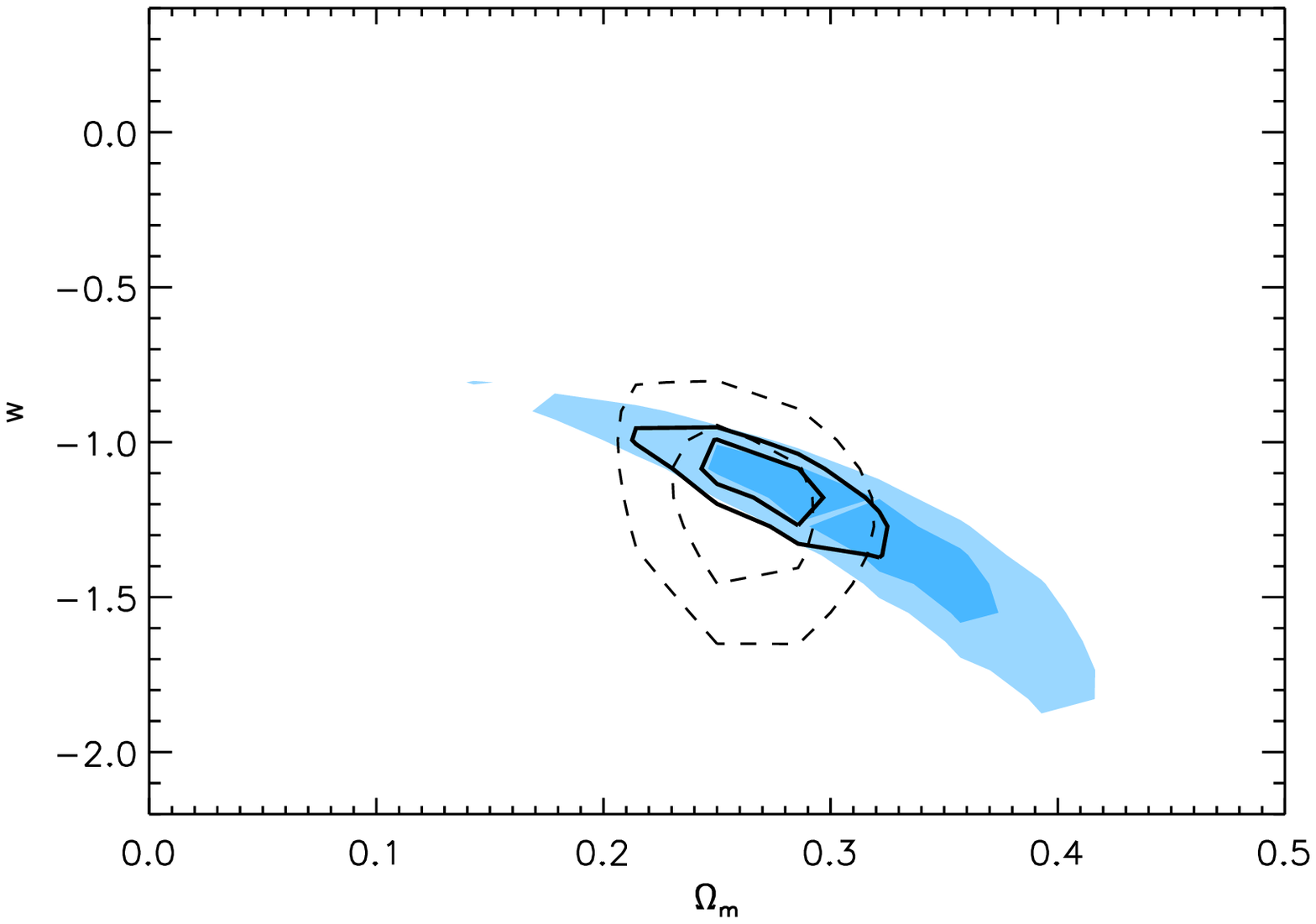}
\caption{{\emph Left:} Current constraints for flat wCDM models, allowing 
for violation of photon number conservation, $k\ne 0$. {\emph Right:} Current 
constraints for flat wCDM models conserving photon-number, $k=0$.}\label{conservpriors_w}
\end{figure*}

In Fig. \ref{conservpriors_w} we show constraints on the dark energy equation
of state $w$ for flat wCDM models, again allowing for $k\in[-5,5]\times 10^{-5}$ 
(left), and enforcing photon number conservation, $k=0$ (right). This again highlights 
the dramatic effect of $k$ on constraints derived from the SN data. On the contrary, 
the effect of $k$ for the H(z) data is insignificant; that is, $H(z)$ alone does not 
significantly constrain $k$ as mentioned above---refer to equation (\ref{rhogen}) 
and the ranges of $w_0$ and $k$.

\begin{figure*}
\begin{center}
\includegraphics[width=8cm]{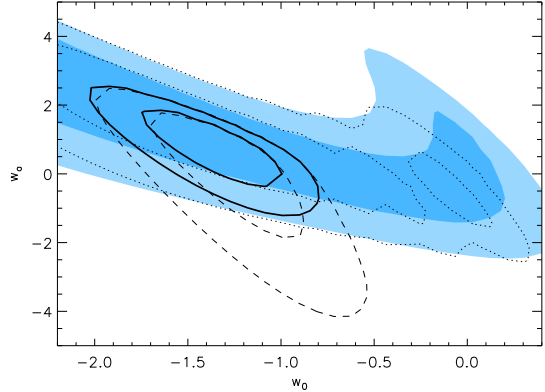} 
\caption{SN+H(z) constraints (solid line contours) on flat CPL-CDM models with a prior on $\Omega_m$ from 
WMAP9+ACT+SPT+BAO+H0. Blue solid contours show 68\% and 95\% confidence levels 
corresponding to the SN data, assuming a flat prior on $k$ in the range $k\in[-5,5]\times 10^{-5}$. 
For comparison, dotted line transparent contours show the corresponding constraint 
assuming a Gaussian prior on $k$ centred at $k=0$ and with $\delta k=10^{-5}$. 
The dashed line contours are for H(z) data and are practically unaffected by the prior on $k$. 
For the SN-only constraints to become comparable to H(z), photon number violations must be 
constrained at a level $\delta k\sim 10^{-6}$.}\label{strongprior}
\end{center}
\end{figure*}

The important effect of photon-number violation on weakening SN constraints, shown 
in Figs. \ref{conservpriors}-\ref{conservpriors_w}, makes clear that, in order to efficiently
exploit current and future SN data for dark energy parameter determinations, one must 
independently constrain photon-number violations, which can be done through independent 
$T(z)$ determinations \cite{Luzzi,Noterdaeme,Tofz}, as discussed above. Alternatively, one may 
try to shrink the SN contours by including information from other cosmological observations as 
priors in the marginalised parameters. Let us for example take the WMAP9+ACT+SPT+BAO+H0 
result $\Omega_m=0.263\pm 0.015$ for the CPL-CDM model (waCDM in the WMAP data tables). 
We repeat our analysis with $\Omega_m$ still in the  interval $[0,1]$ but this time assuming a 
Gaussian prior (centred at 0.263 and with standard deviation 0.015) when marginalising over 
$\Omega_m$. In this case, we also remove the corresponding BAO datapoints from our H(z) 
sample (the 3 WiggleZ datapoints at $z=0.44, 0.60$ and 0.73). The resulting constraint on 
the $w_0 - w_a$ plane is shown in Fig. \ref{strongprior}. The stronger prior on $\Omega_m$ 
now disfavours the region with large $\Omega_m$ and negative $k$, thus leading to an extension 
of the 68\% SN contour towards the region with $\{w_0\!\sim\!-1, \Omega_m\!\sim\!0.3, k\!\sim\!0\}$ 
(also refer to Fig.~\ref{conservpriors}). However, the degeneracy is not broken, and the region 
around $w_0\sim 0$ is still allowed by the SN data. The overall constraint on $w_0$ is still controlled 
by the H(z) data, even though the bound on $w_a$ is now much stronger.  

In the same figure, we also show the corresponding SN contours that one would obtain 
if $k$ was constrained at the $10^{-5}$ level (dotted lines). The SN contours shrink 
a little, but $w_0\sim 0$ is still allowed, and the total constraint is again controlled by 
the H(z) data. For the Union2.1 SN dataset to become competitive with the current H(z) 
determinations, $k$ must be independently constrained at the $\sim\! 10^{-6}$ level, 
resulting in a SN contour similar to that of Fig~\ref{conservpriors}, bottom right. This is at 
a level attainable using current Sunyaev-Zel'dovich T(z) measurements: namely the stacked 
Planck SZ clusters of \cite{Hurier13} and the constraints from SPT clusters \cite{Saro13}. 
We will present a detailed T(z) analysis based both on SZ clusters and on atomic and 
molecular absorption lines in a follow-up publication. Such T(z) constraints will therefore 
provide an independent means of
breaking the degeneracy between dark energy parameters and photon number non-conservation, 
subject to completely different systematics. Currently, this degeneracy can only be broken by 
combining SN constraints with a $k$-independent dataset like $H(z)$, as we have demonstrated, 
and it is important to be able to also break the degeneracy by directly constraining the photon violation 
parameter $k$. Similarly, if one is instead interested in constraining photon-number violations in the 
context of evolving dark energy cosmologies (rather than constraining dark energy in a way that 
accounts for photon-number violations), then T(z) measurements offer a more direct way, 
complementary to standard (but indirect) distance duality analyses. Further, expected improvements 
in SN and H(z)/BAO data in the next decade or so may not be able to reach the desired 
$\delta k\lesssim\! 10^{-6}$ level in the context of evolving dark energy models using distance-duality 
methods, as we discuss in the next section.        

\section{Constraints from future datasets}

Let us now proceed to present forecasts for future datasets. We will be particularly interested in
studying the impact of the Euclid mission, and will start by considering one of its probes, BAO from the Euclid wide survey,
which is expected to cover $\sim$15000 deg$^2$ of extragalactic sky down to a redshift of order 2. Assuming the same 
conservative priors as in Fig. \ref{conservpriors}, we repeat the above analysis 
with simulated data for Euclid BAO and a SNAP-like SN mission. BAO radial 
distance errors have been estimated using the code developed by Seo \& 
Eisenstein \cite{SeoEisen} adapted for Euclid estimated parameters \cite{Euclid}. For SN we 
follow \cite{DETaskForce} for a Dark Energy Task Force Stage IV SN mission.  

\begin{figure*}
\includegraphics[width=8cm]{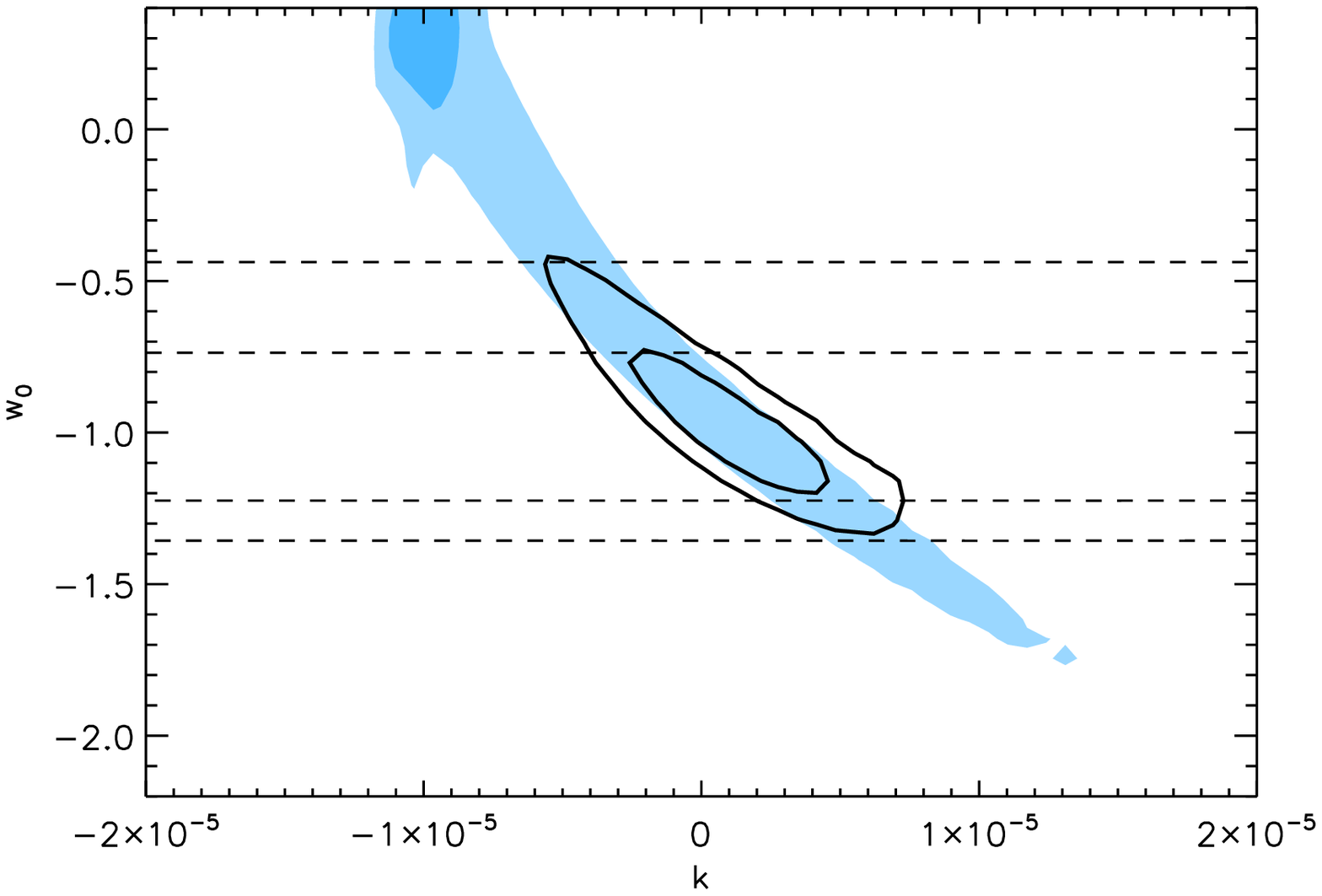} 
\includegraphics[width=8cm]{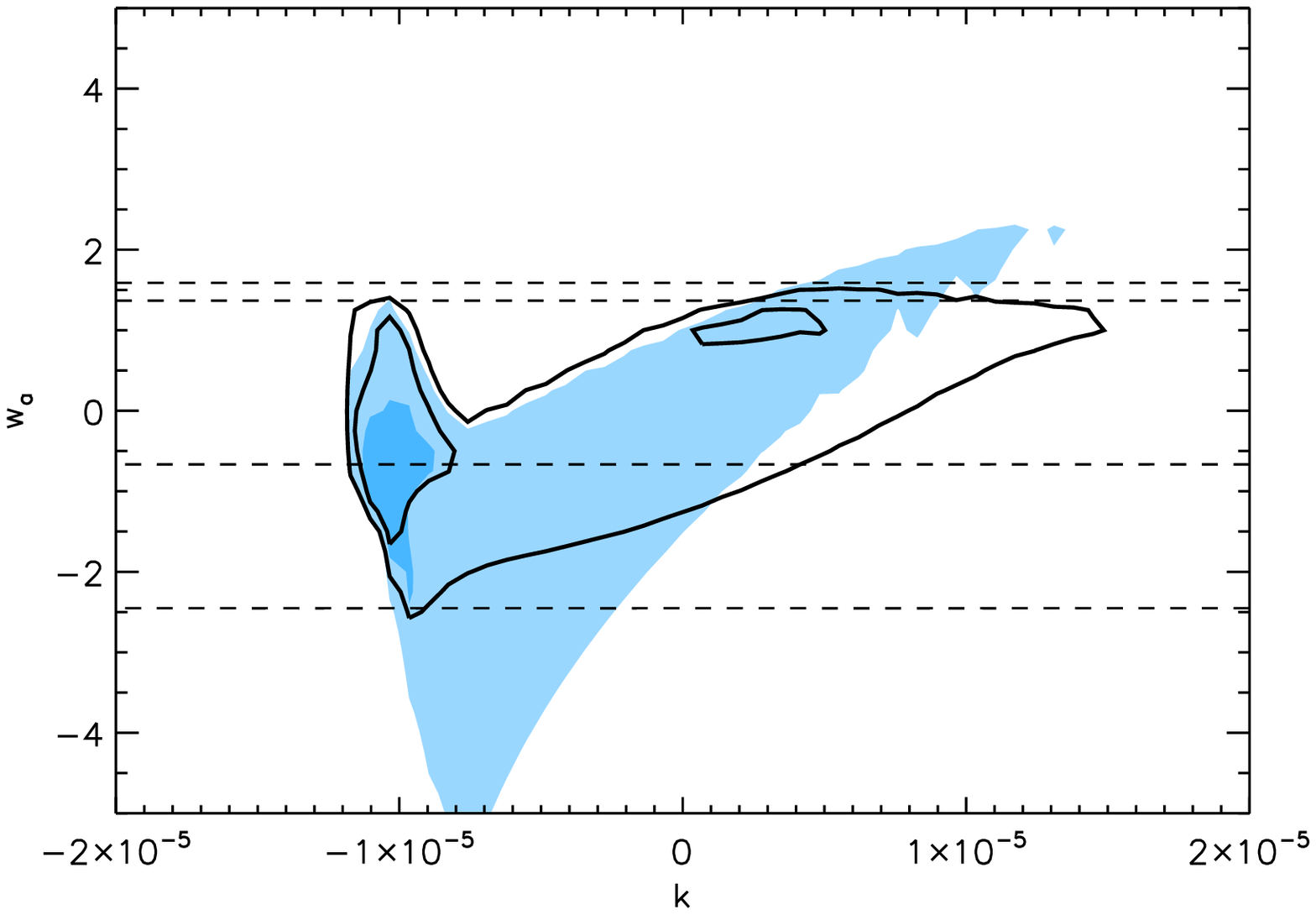}
\includegraphics[width=8cm]{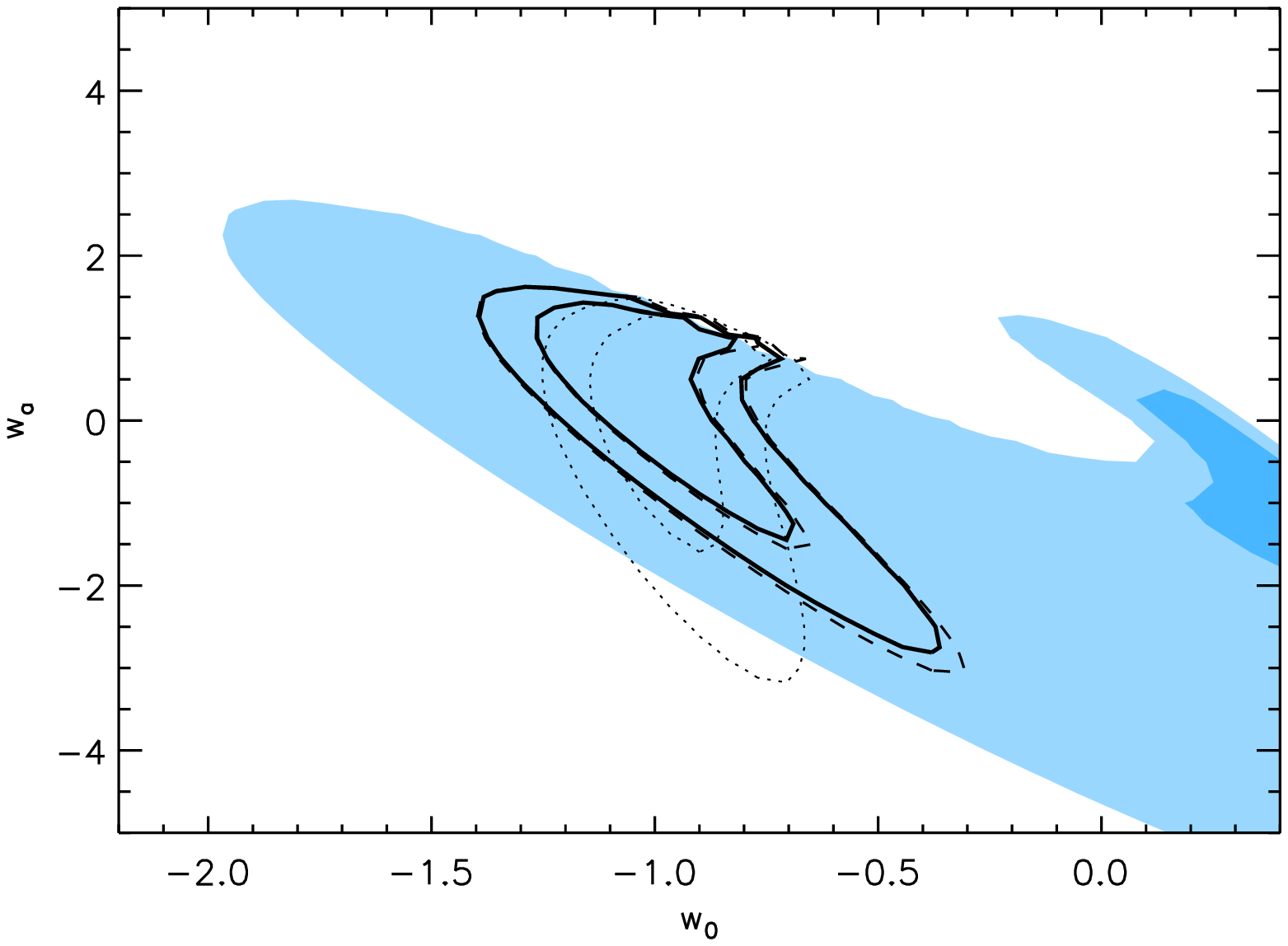} 
\includegraphics[width=8cm]{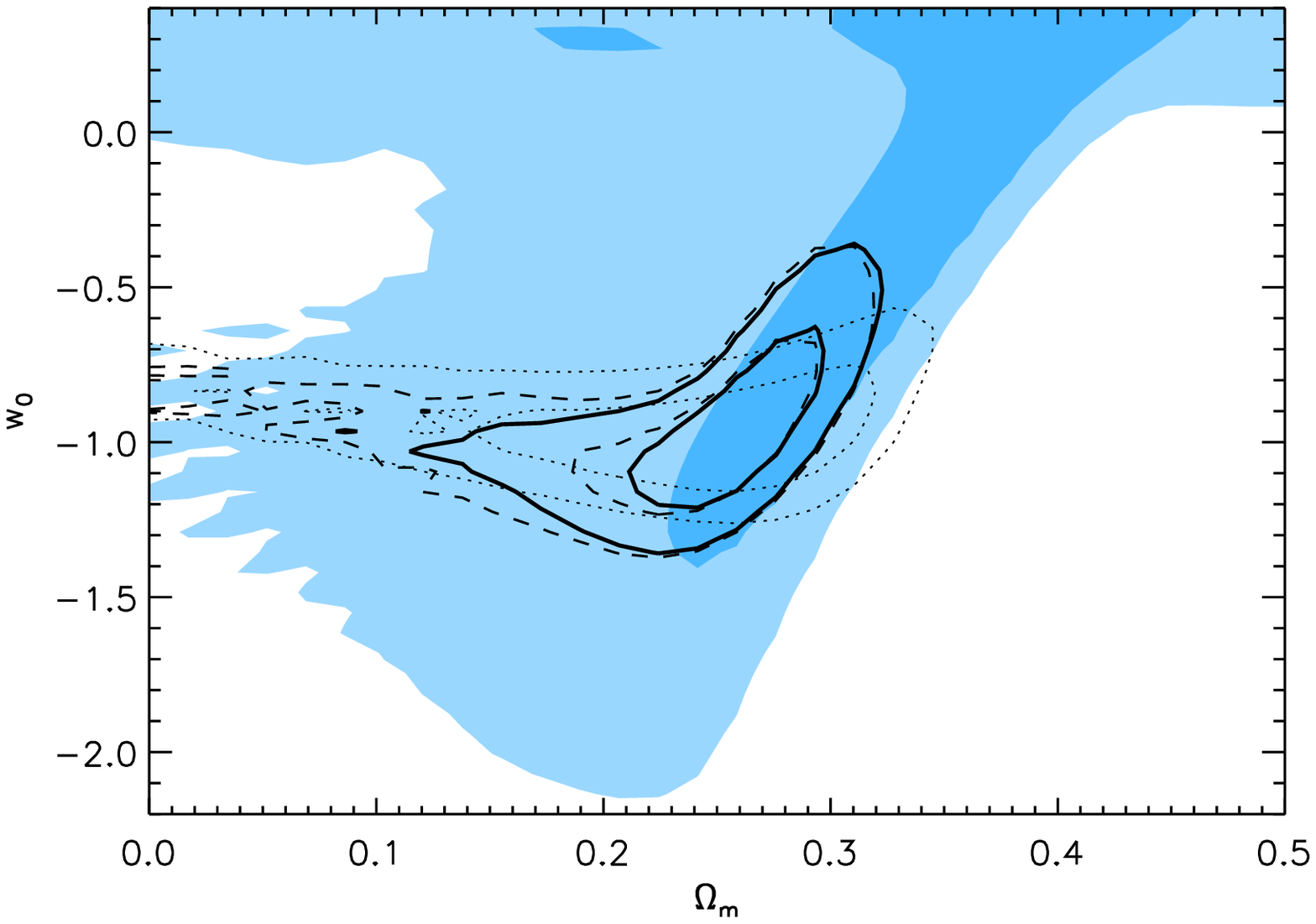}  
\includegraphics[width=8cm]{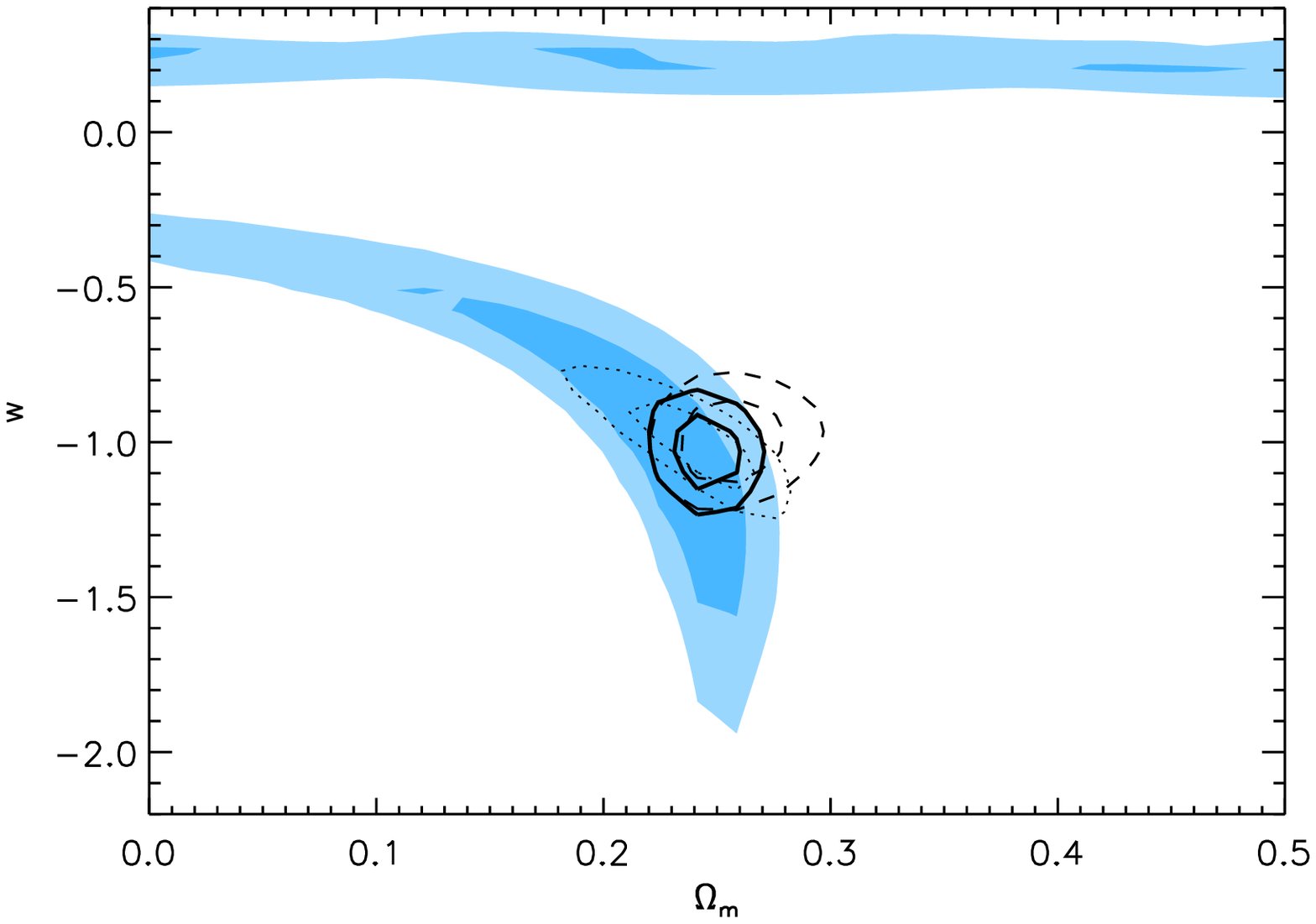} 
\includegraphics[width=8cm]{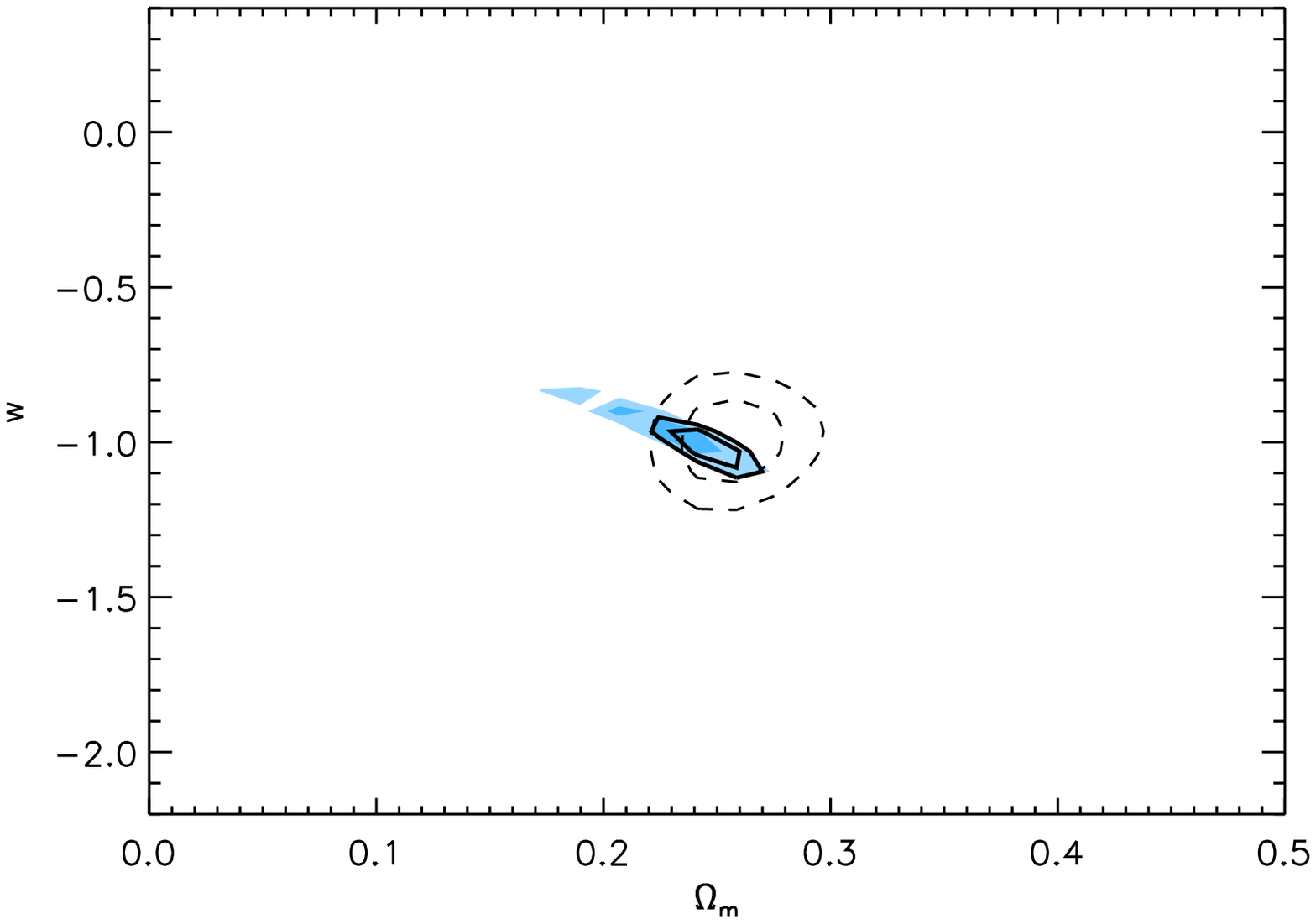}
\caption{ 
Forecast 68\% and 95\% likelihood contours for SN (filled blue), H(z) (dashed line 
transparent) and combined SN+H(z) (solid line transparent), considering Euclid BAO 
and a SNAP-like SN mission. The dotted transparent contours show the corresponding 
SN likelihood contours, assuming $k$ can be constrained at the $10^{-6}$ level.   
\emph{Top and Middle Panels:} Forecasts for flat CPL-CDM models, for the conservative priors 
of Fig. \ref{conservpriors}. 
\emph{Bottom Panel:} For comparison we show the corresponding constraints 
for a wCDM model for $k\in[-5,5]\times 10^{-5}$ (left) and for $k=0$ (right). On the 
left plot, note that the horizontal band around $w\gtrsim 0$ for the SN data corresponds to a region 
with $k\simeq -10^{-5}$ (cf. top plots), and it disappears when $k$ is constrained at the $10^{-6}$ 
level.}\label{BAOSNforec}
\end{figure*}

The relevant constraints are shown in Fig. \ref{BAOSNforec}. As before, 
we show 68\% and 95\% likelihood contours for the SN data in blue (filled contours), 
while the transparent dashed line contours are for BAO H(z) data. The combined constraints are 
shown as solid line, transparent contours.  Note in Fig.~\ref{BAOSNforec} that the constraint 
on $w_a$ starts to become interesting with Euclid BAO+SNAP-like SN, even allowing for photon 
number non-conservation. However, in the absence of an independent constraint on $k$, much 
of the observed SN dimming can be explained by photon conversion to scalar particles, so the 
SN contours grow (compared to the case $k=0$) and the total constraint is dominated by $H(z)$.

To highlight this point, we also show in the middle and bottom left panels (dotted, transparent contours) 
the SN constraints achievable if the photon number violation parameter, $k$, could be constrained at 
the level $10^{-6}$. This can be seen to have an important effect on the joint contours, as the SN-only 
constraints become competitive to the H(z)-only ones. Note, in particular, that the horizontal 
band around $w\gtrsim 0$ (bottom left plot) for the SN data disappears, as it corresponds to a region 
with $k\simeq -10^{-5}$ (cf. top plots). The main message arising from this analysis is that constraining 
$k$ at a level $\lesssim\! 10^{-6}$ will rule out this region, shrinking the SN contours by a factor of a few 
and improving the joint SN+H(z) constraint by a factor of $\sim 2$ in the context of CPL-CDM models 
(Fig.~\ref{BAOSNforec}, middle left plot). We will discuss current and future constraints on $k$ from 
T(z) measurements, and their quantitative effect on dark energy parameter bounds, in a follow-up publication. 
In the rest of this section we will examine the ability of future BAO and SN surveys to jointly constrain dark energy 
parameters and $k$ in the absence of such an independent bound on $k$.  

\begin{figure*}
\includegraphics[width=8cm]{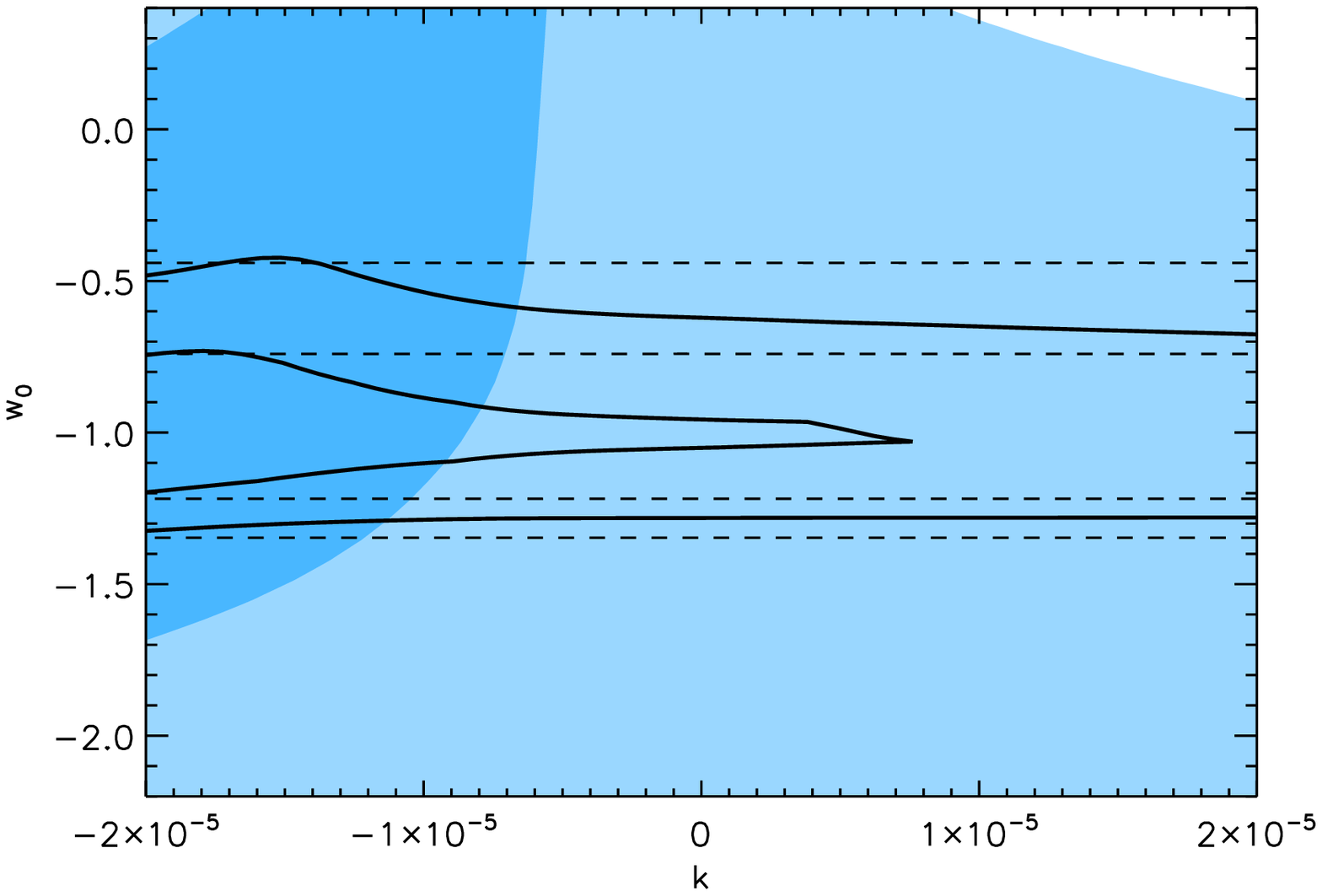} 
\includegraphics[width=8cm]{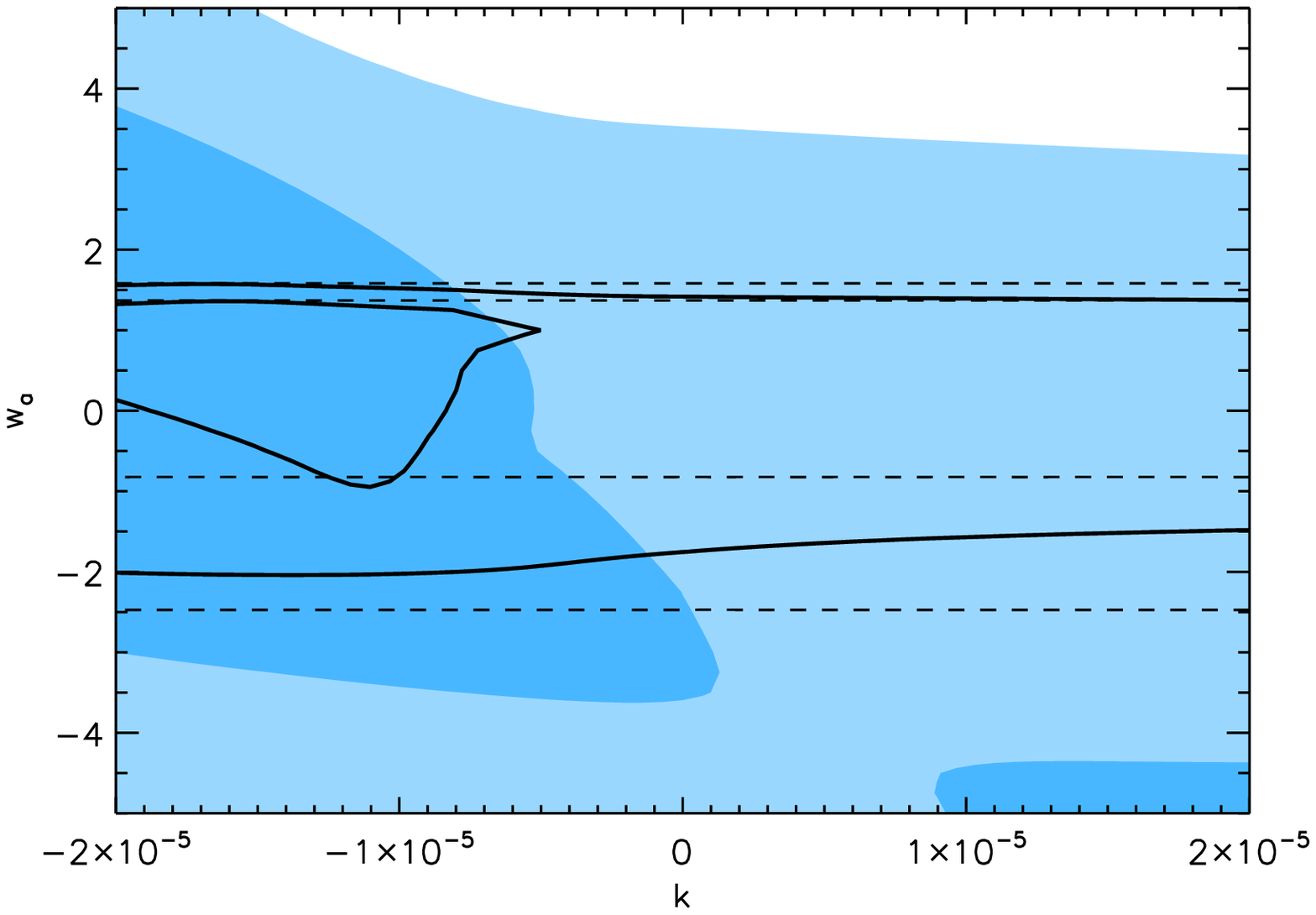}
\includegraphics[width=8cm]{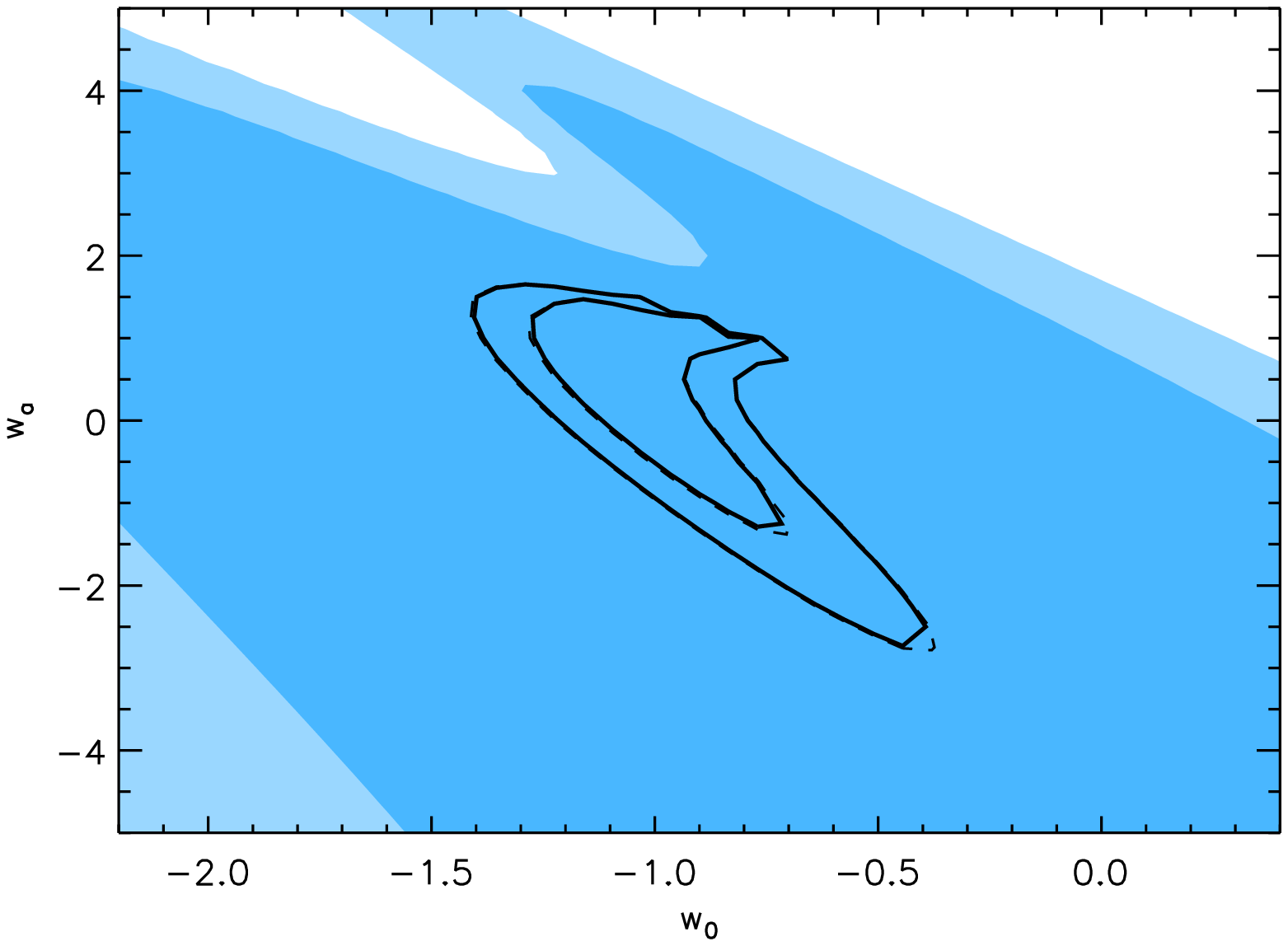}
\includegraphics[width=8cm]{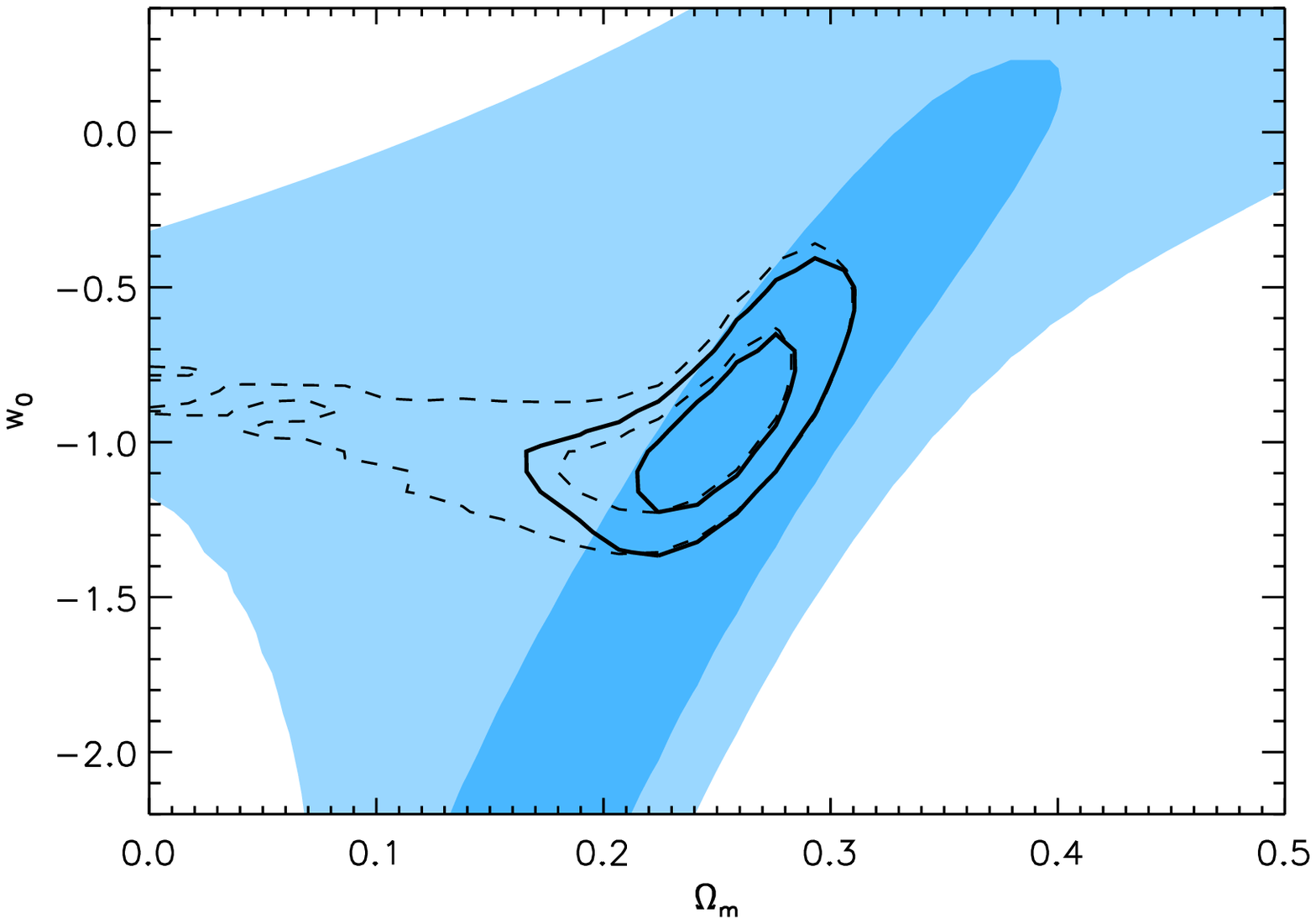}   
\includegraphics[width=8cm]{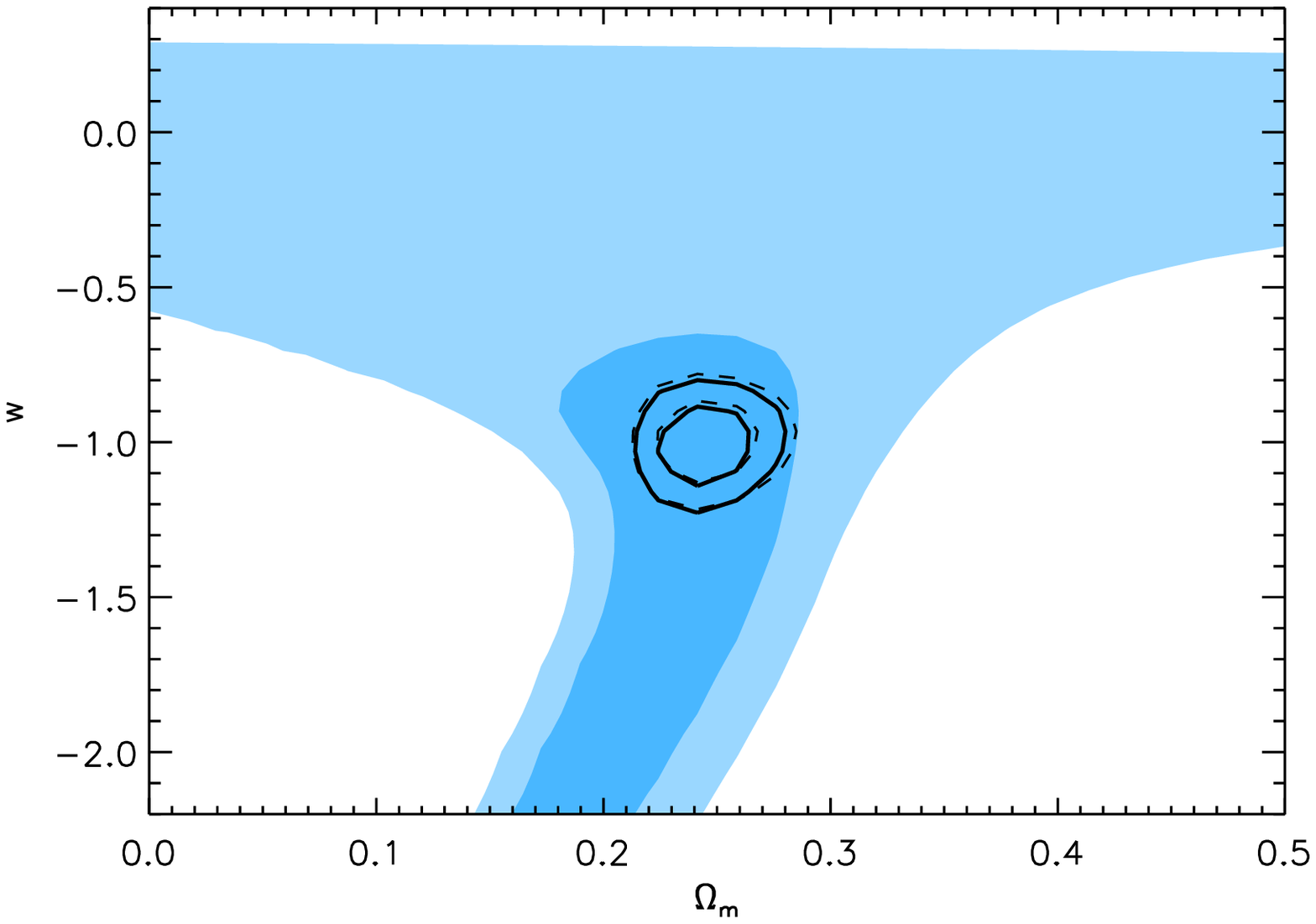} 
\includegraphics[width=8cm]{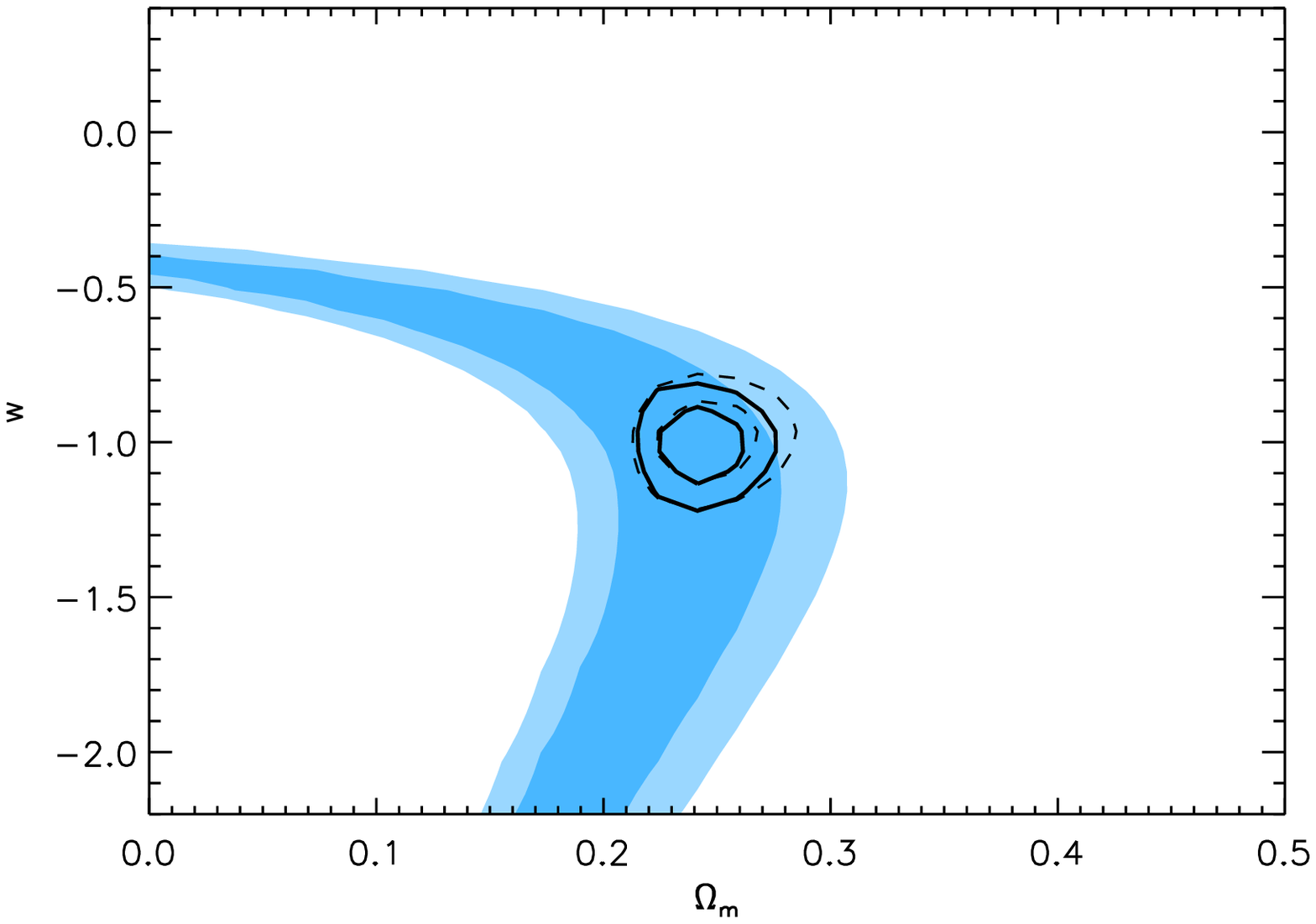}
\caption{Forecasts for Euclid BAO (wide survey) + Euclid AAA SN survey. The plots shown 
are as in Fig.~\ref{BAOSNforec}. While Euclid alone cannot strongly constrain $k$ (violations 
of photon number conservation) it will still provide useful constraints on dark energy parameters, 
especially within the wCDM model. However, the joint constraints (solid transparent contours) are 
determined predominantly by the BAO data (dashed transparent contours), while the SN-only 
bounds (blue filled contours) are weak. Therefore, these constraints become much stronger 
when a low-redshift SN sample is a also included, cf. Fig.~\ref{EuclidSNAP}.}\label{EuclidOnly}
\end{figure*}

\begin{figure*}
\includegraphics[width=8cm]{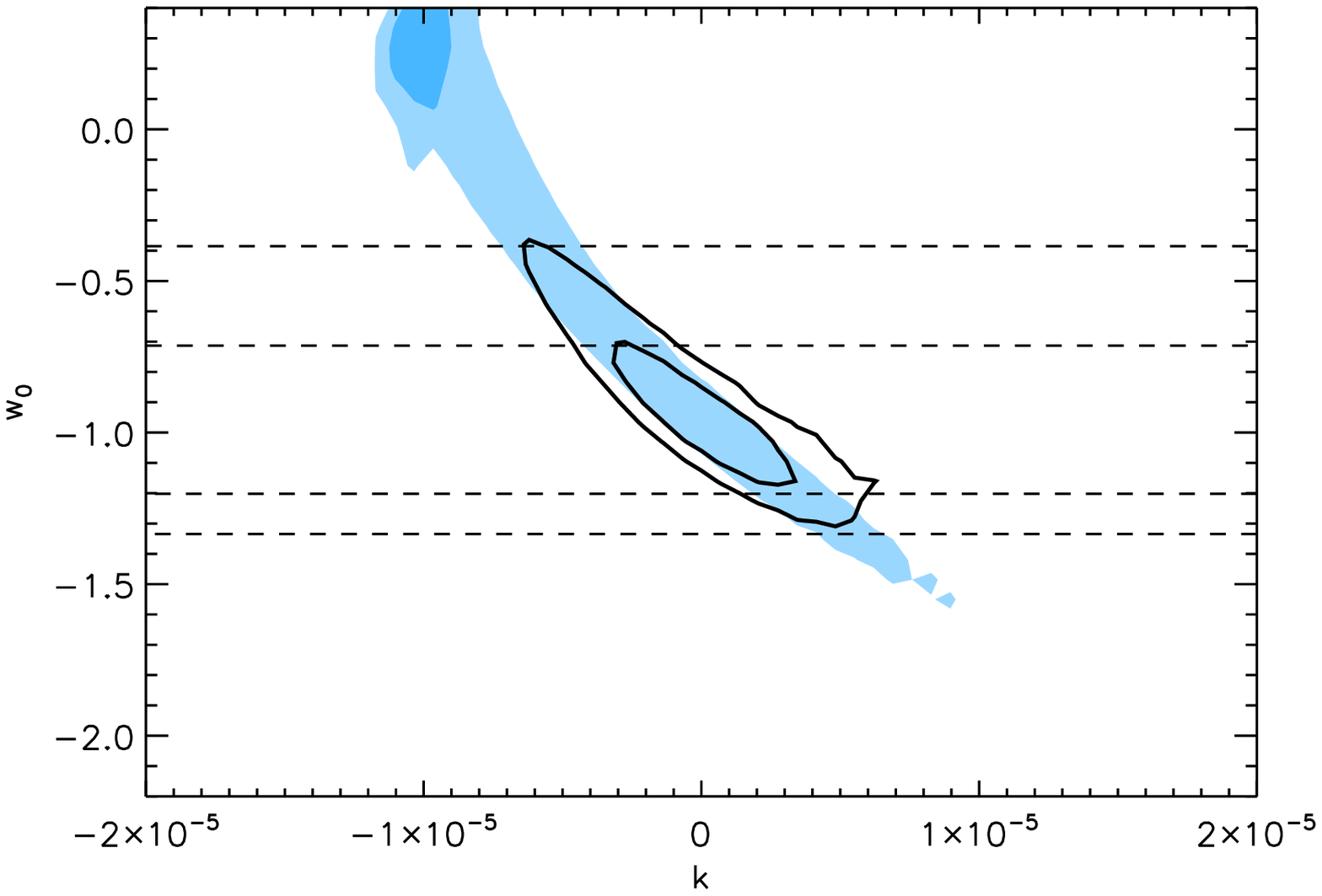} 
\includegraphics[width=8cm]{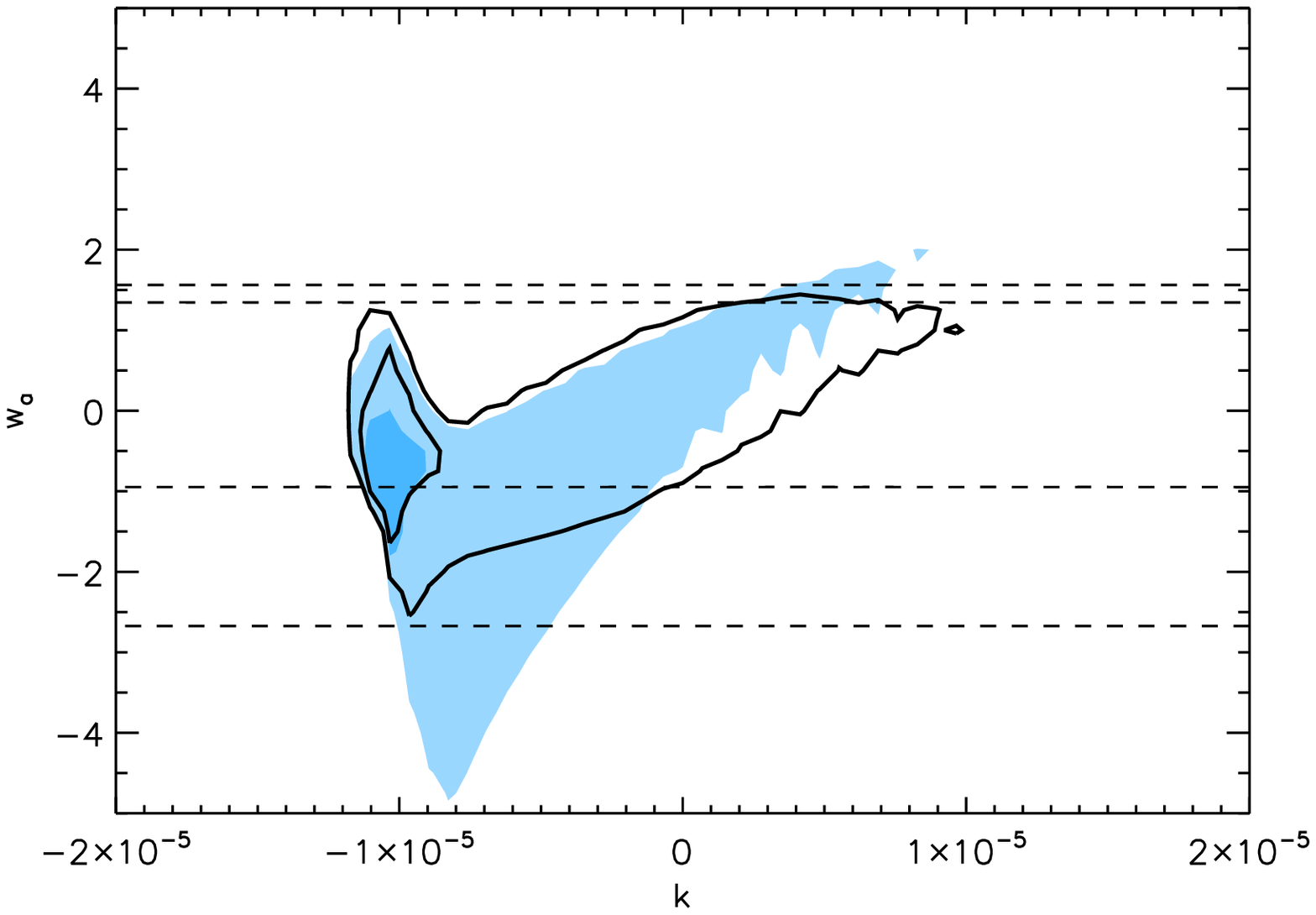}
\includegraphics[width=8cm]{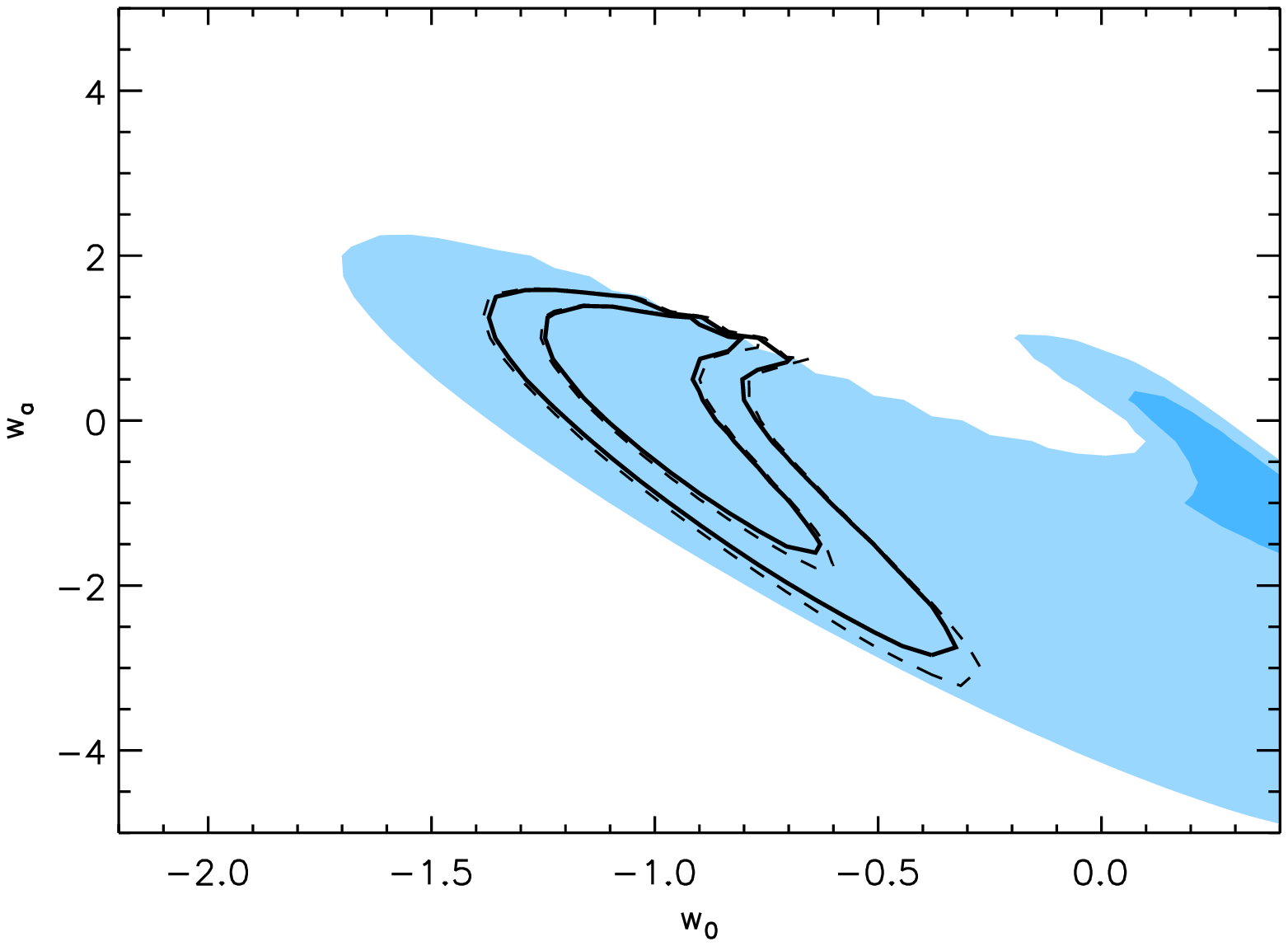}
\includegraphics[width=8cm]{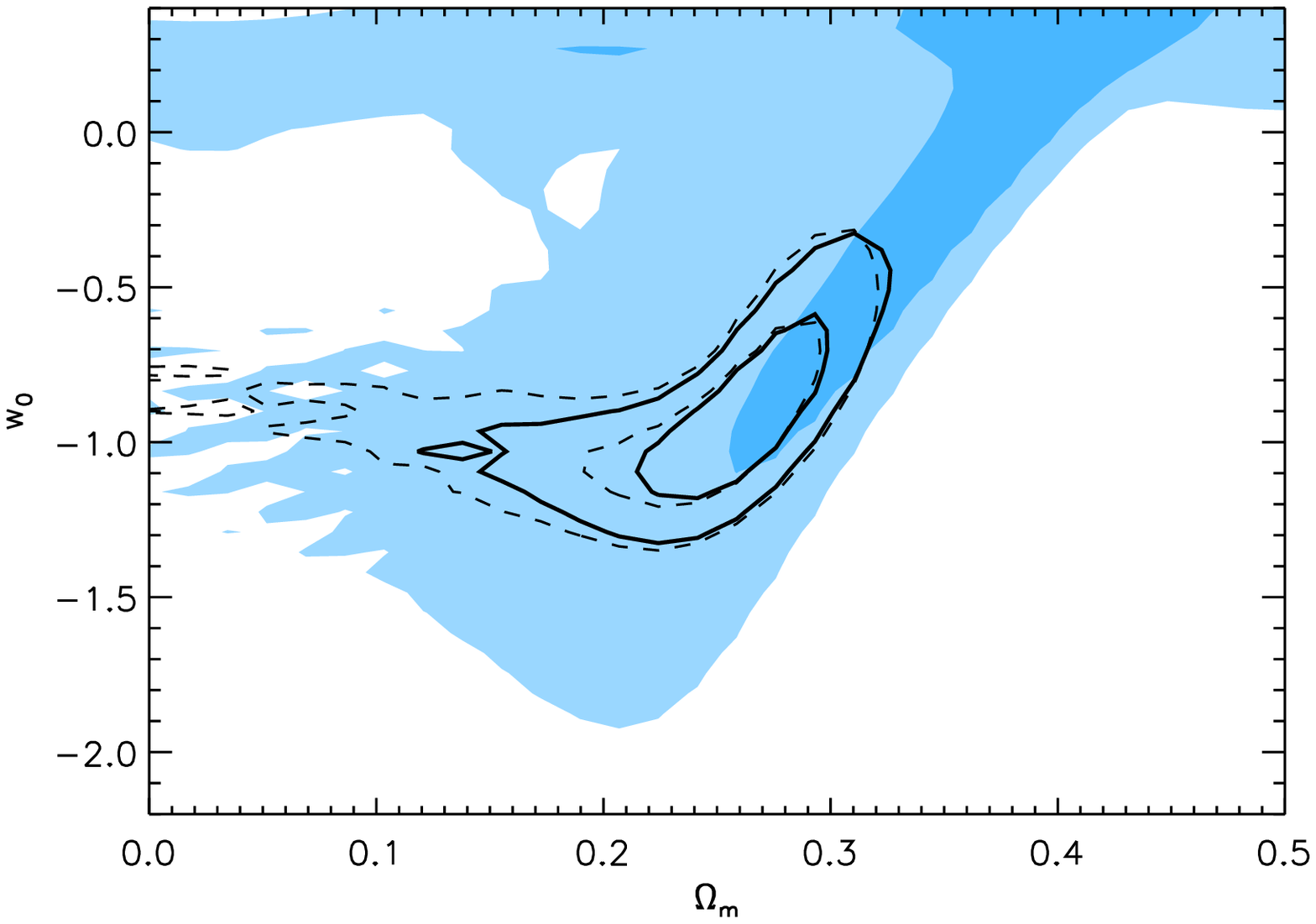}   
\includegraphics[width=8cm]{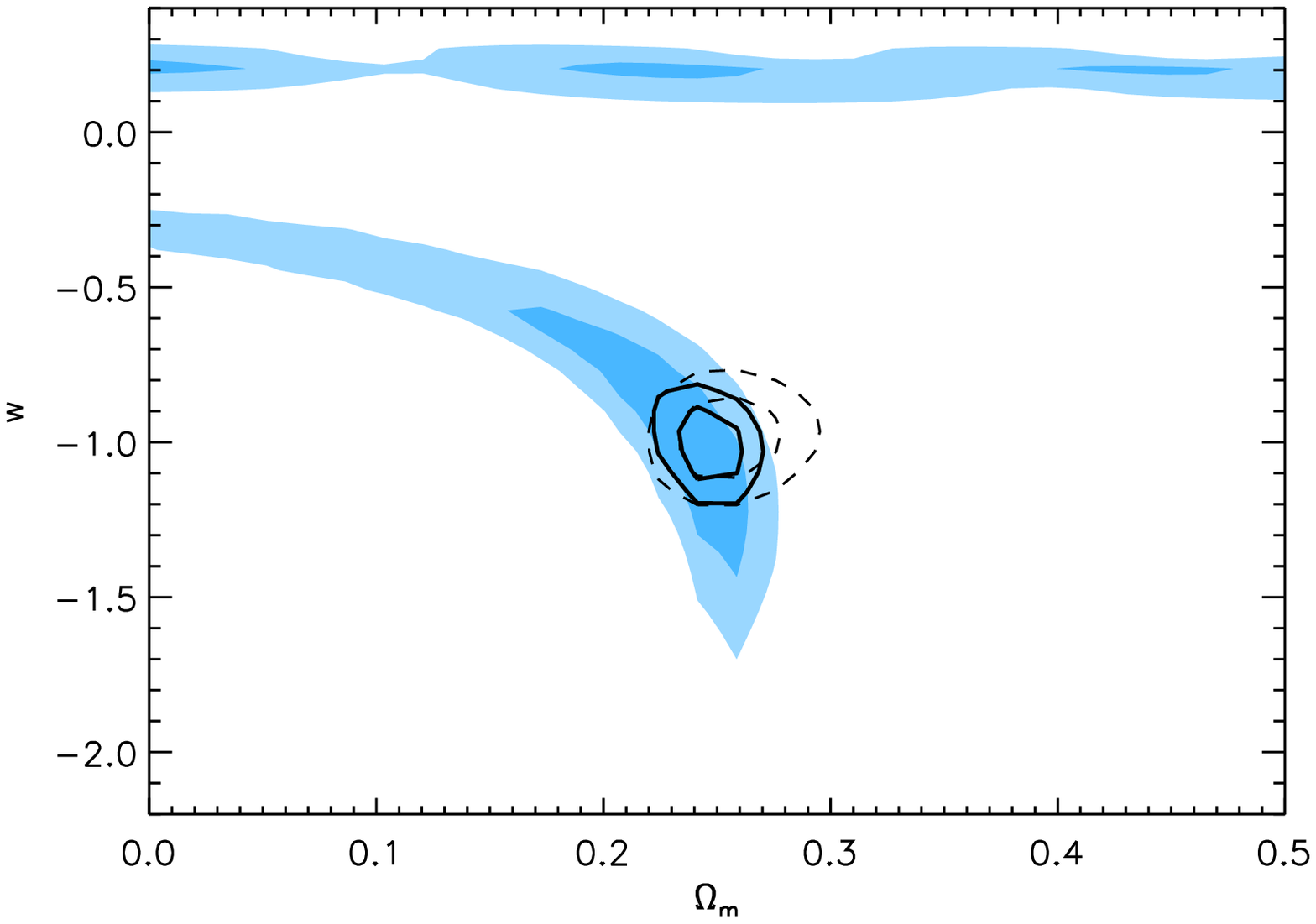} 
\includegraphics[width=8cm]{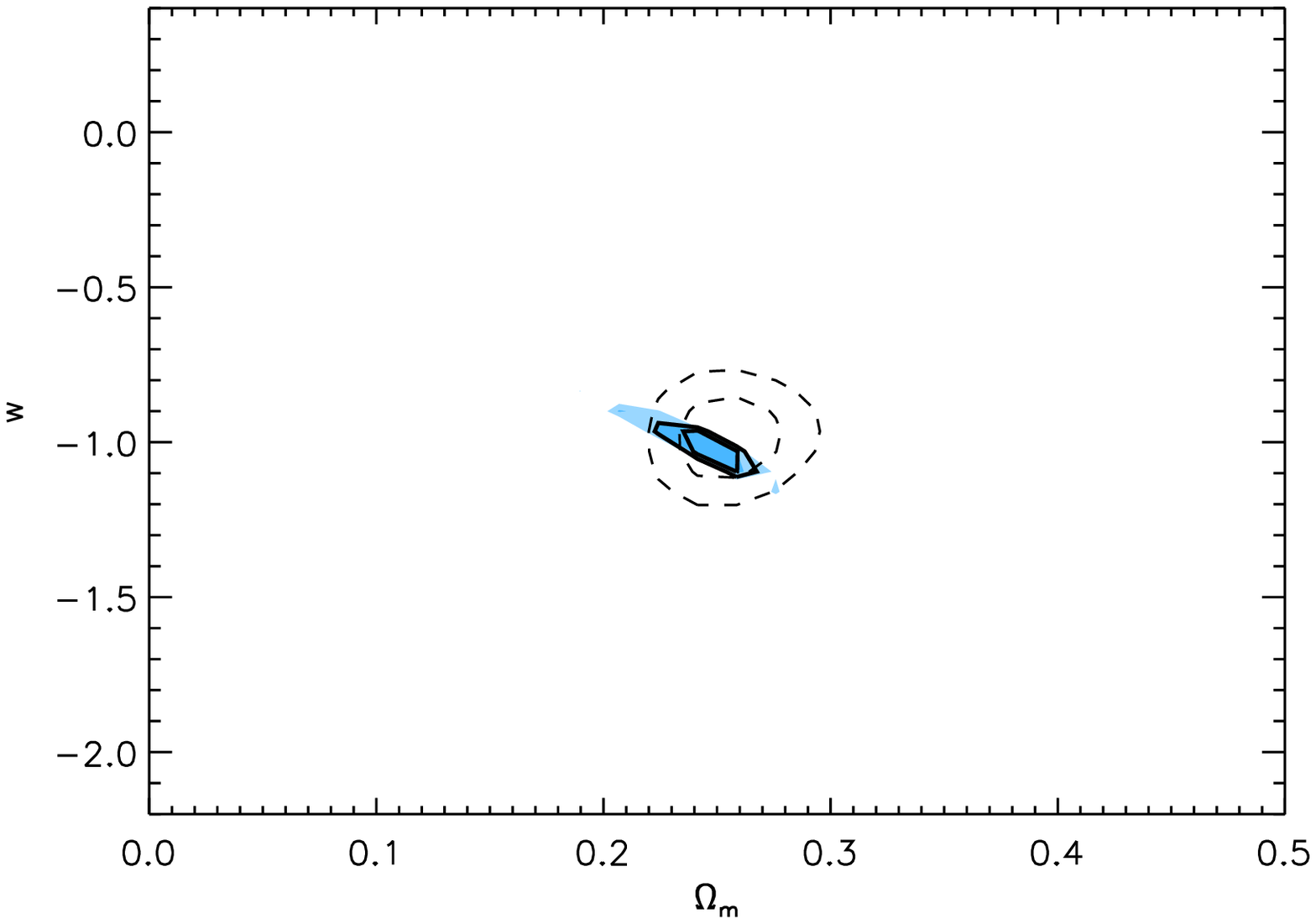}
\caption{Forecast for Euclid (SN+BAO) combined with a low-redshift, SNAP-like sample. The 
combined SN constraints are now stronger (cf. Fig.~\ref{EuclidOnly}), improving significantly 
the joint constraints.}\label{EuclidSNAP}
\end{figure*}

It has been proposed that Euclid carries out a dedicated SN survey, which could yield 
up to a few thousand SNeIa up to redshift 1.5 \cite{Hook}. This makes Euclid an ideal instrument to constrain the models we are studying, capable of delivering both radial/angular diameter distance measurements and luminosity distances, 
and thus minimising systematics. Based on the recently studied 6-month `AAA' Euclid survey \cite{Astier},
one can expect more than 1700 SNe Ia in the redshift range $0.75<z<1.55$. We adopt the assumptions in this study
and repeat our forecast analysis, now using only Euclid for both BAO and Supernovae. (We neglect the correlation between errors in
different redshift bins; the effects of doing this are expected to be relatively small, given the other approximations we are also making.) Our results are shown in Fig.~\ref{EuclidOnly} showing that Euclid can, even on its own, provide useful 
constraints on Dark Energy allowing for photon number non-conservation, especially 
for wCDM models. Note however, that the SN-only constraints (blue filled contours) are 
weak and the joint constraint is dominated by the BAO H(z). Further, photon-number violations, 
parametrised by $k$, cannot be constrained in this prior range by Euclid alone. Naturally, these 
constraints become much stronger by combining the Euclid SN with a low-redshift sample, 
e.g. from a SNAP-like mission (Fig.~\ref{EuclidSNAP}).    

\begin{figure*}
\includegraphics[width=8cm]{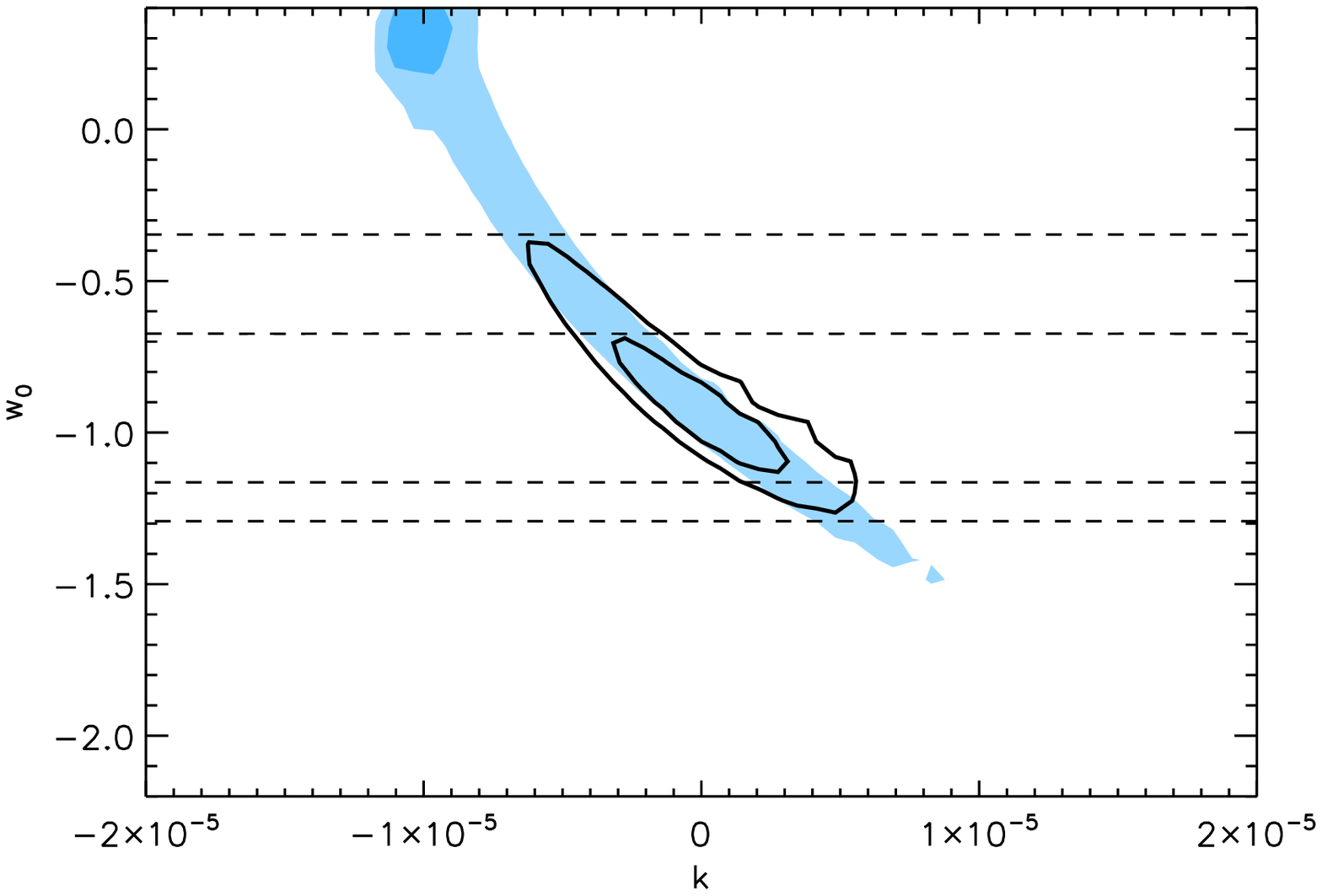} 
\includegraphics[width=8cm]{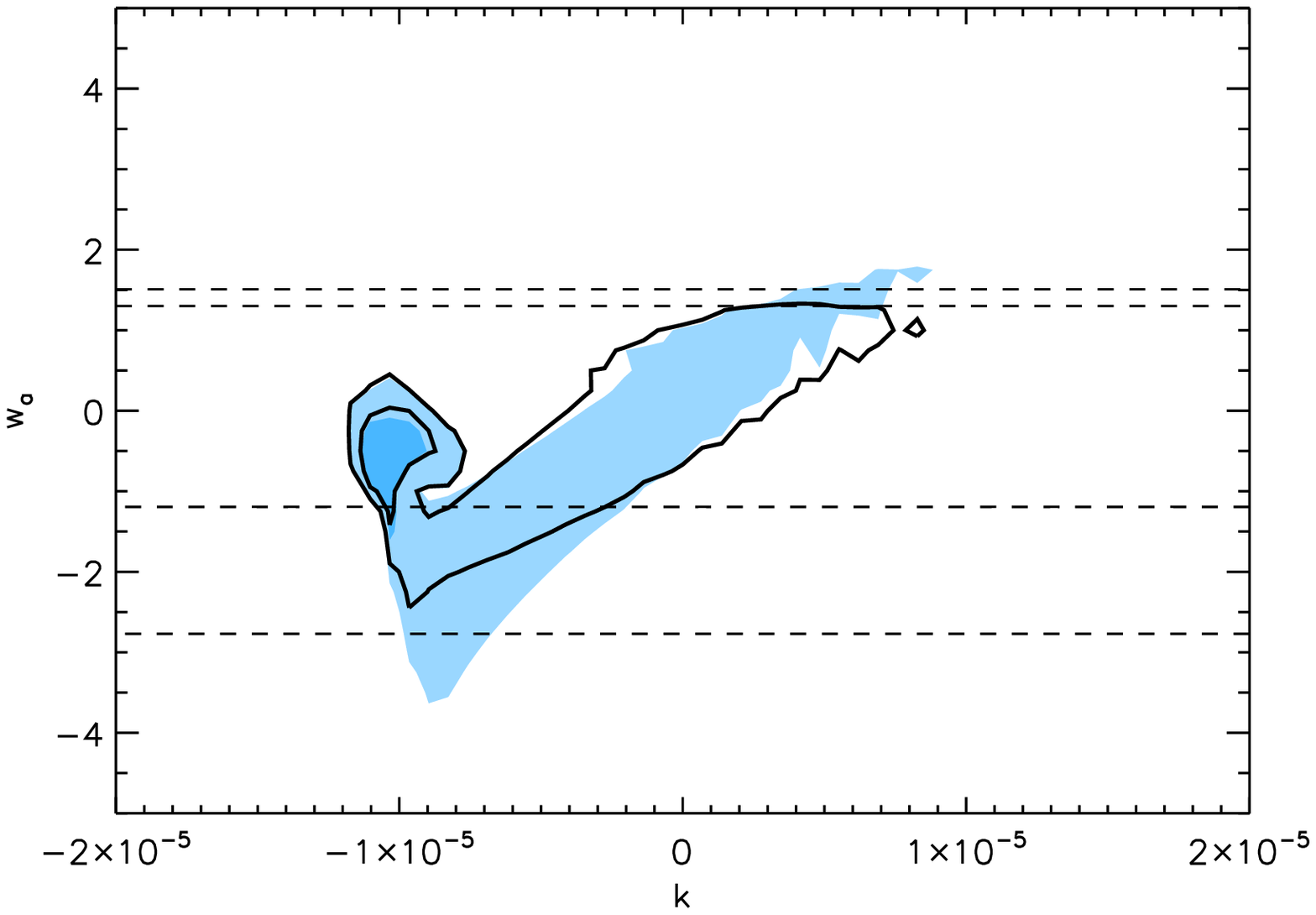} 
\includegraphics[width=8cm]{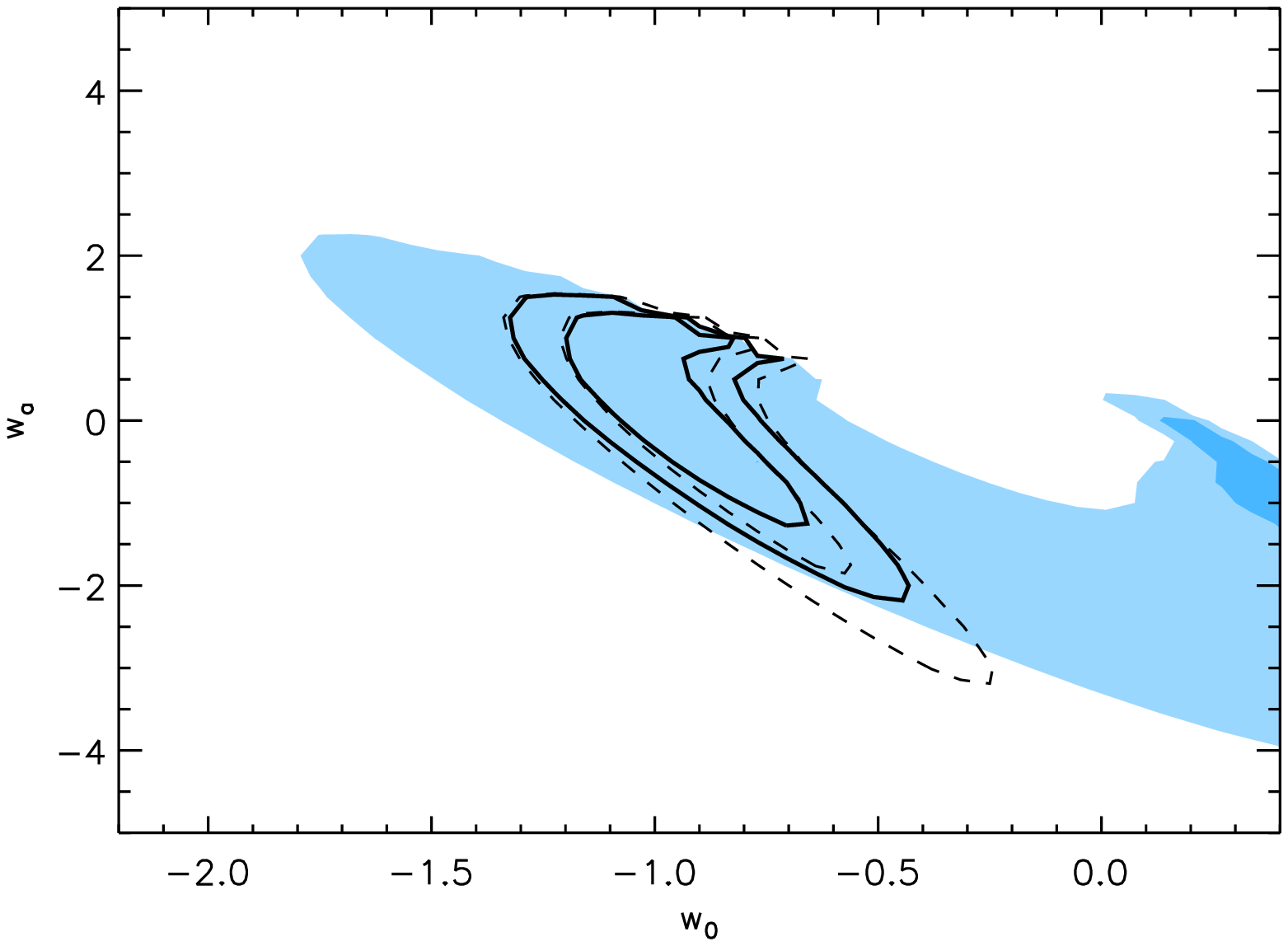}
\includegraphics[width=8cm]{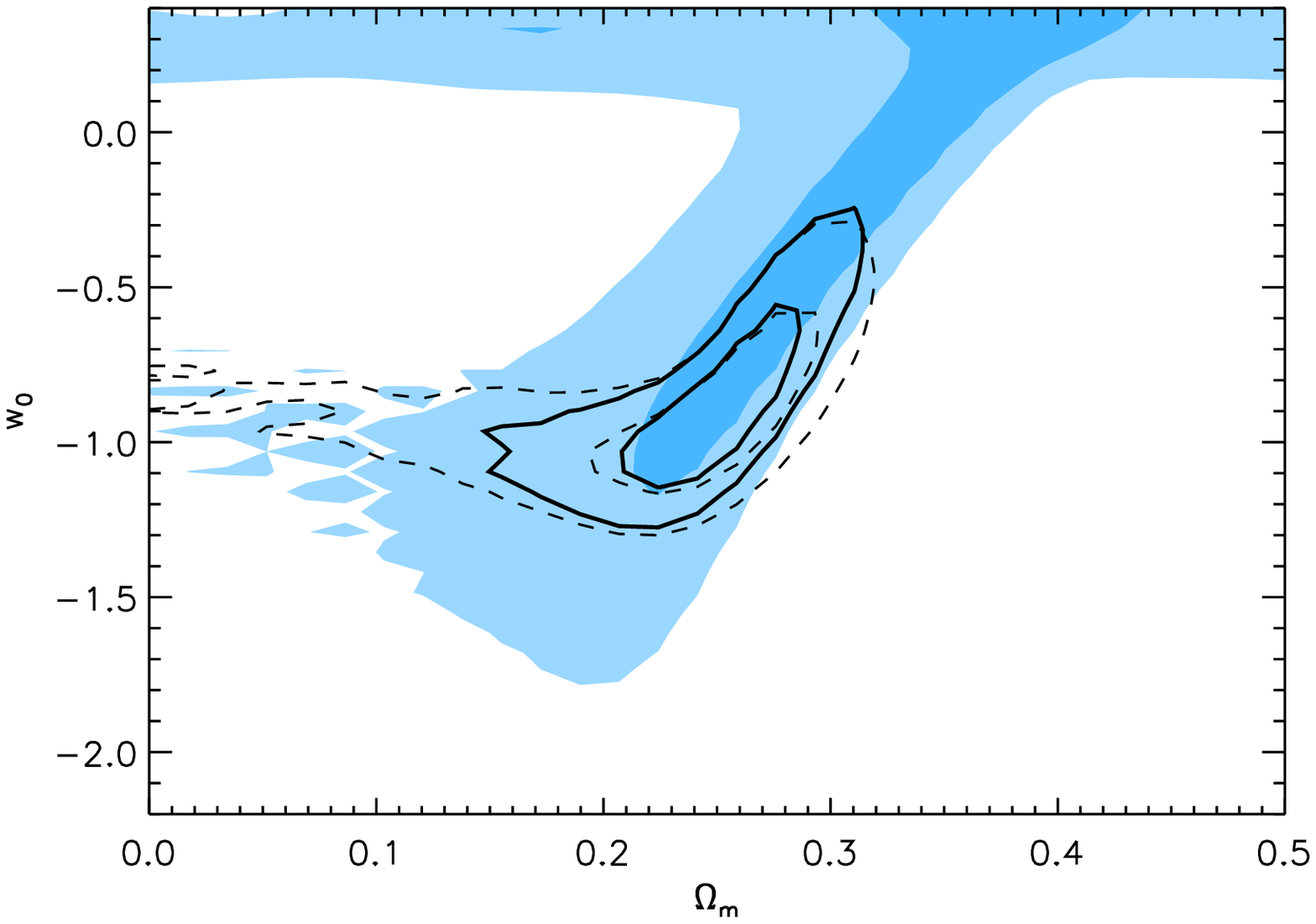} 
\includegraphics[width=8cm]{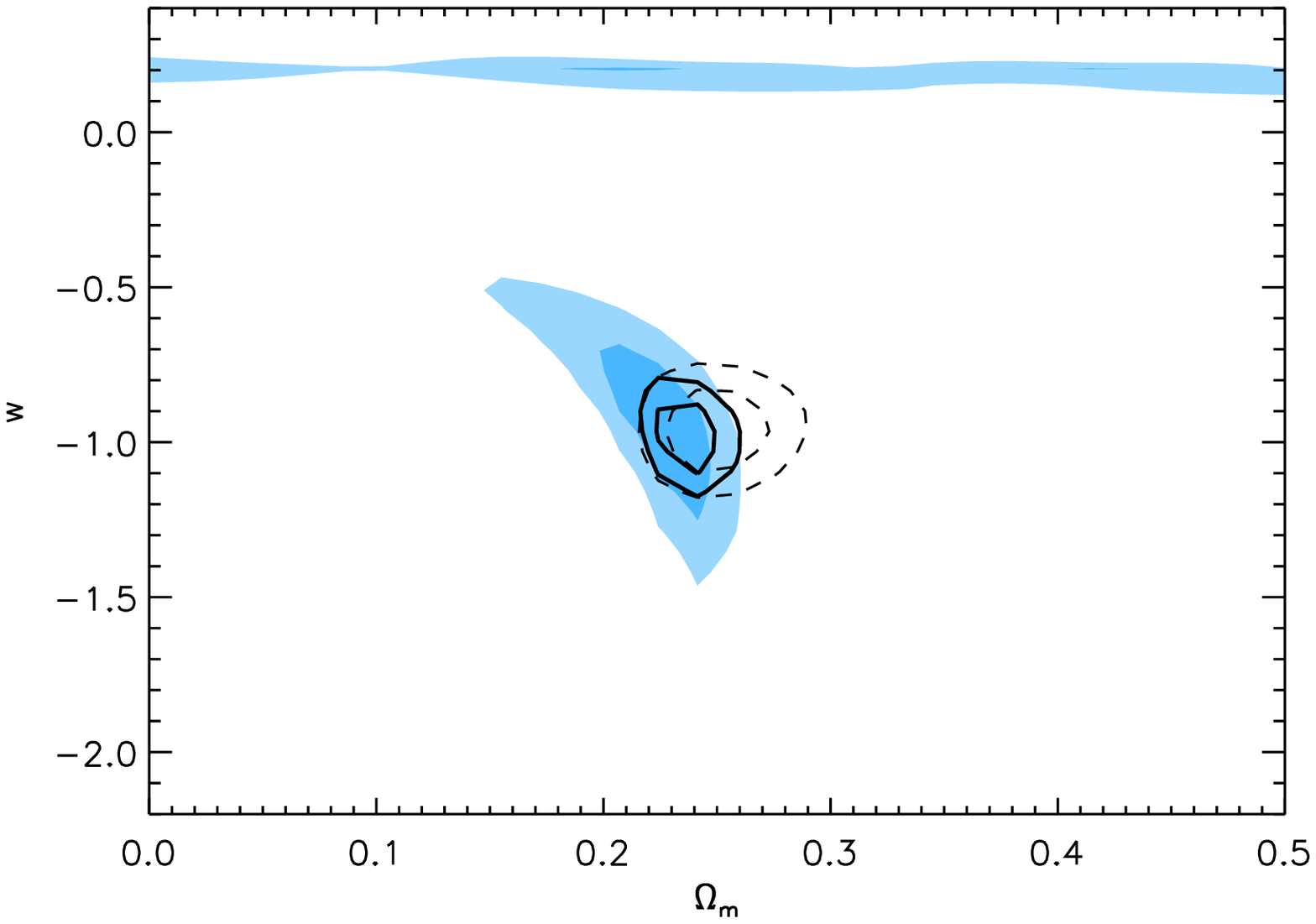} 
\includegraphics[width=8cm]{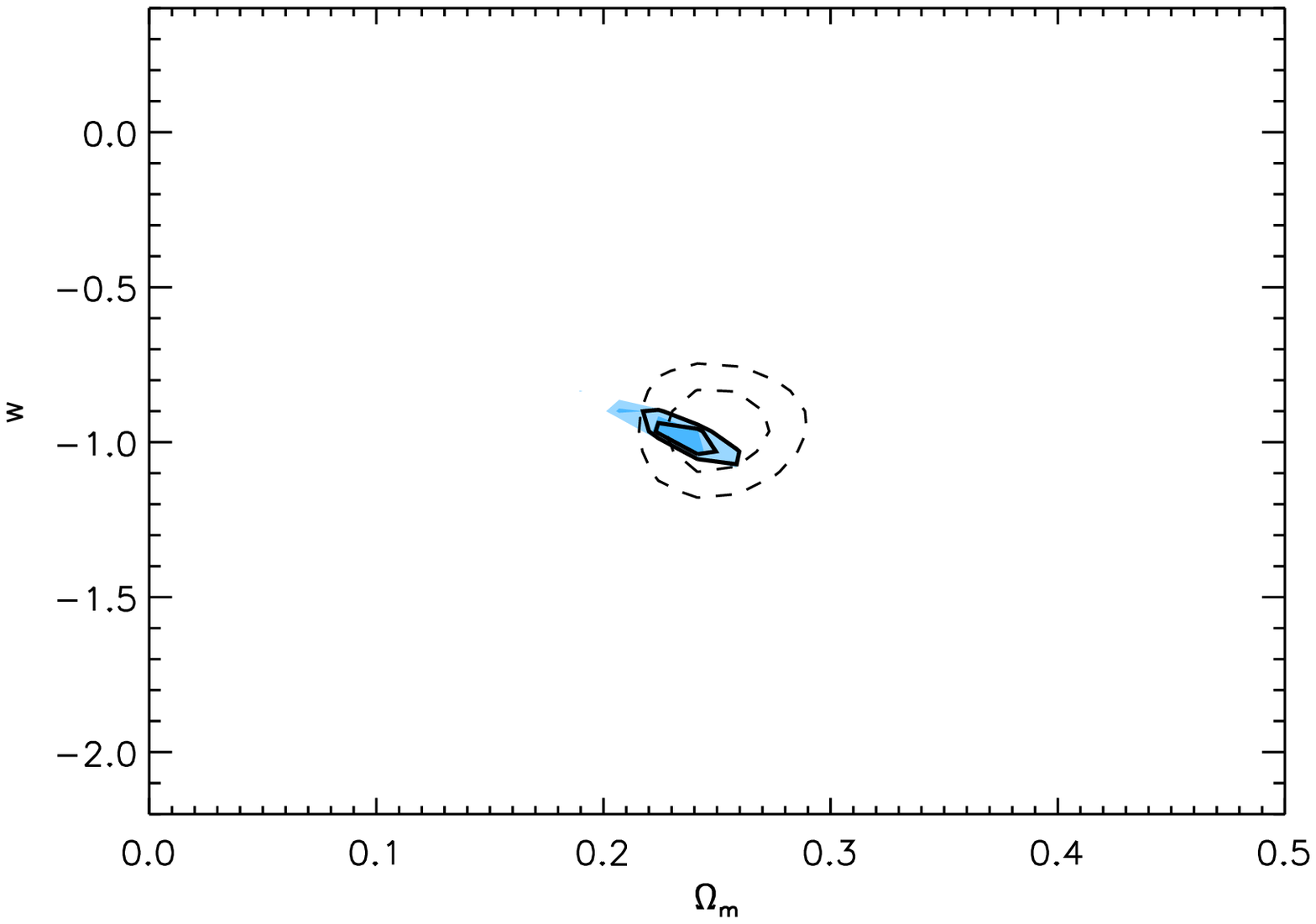}
\caption{As above, but now showing forecasts for E-ELT combined with a SNAP-like SN 
mission and Euclid BAO. The additional $\sim 50$ SN in the range $1<z<5$ can lead to 
a significant improvement of the SN constraints.}\label{EELTSNAPBAO}
\end{figure*}

\begin{figure*} 
\includegraphics[width=8cm]{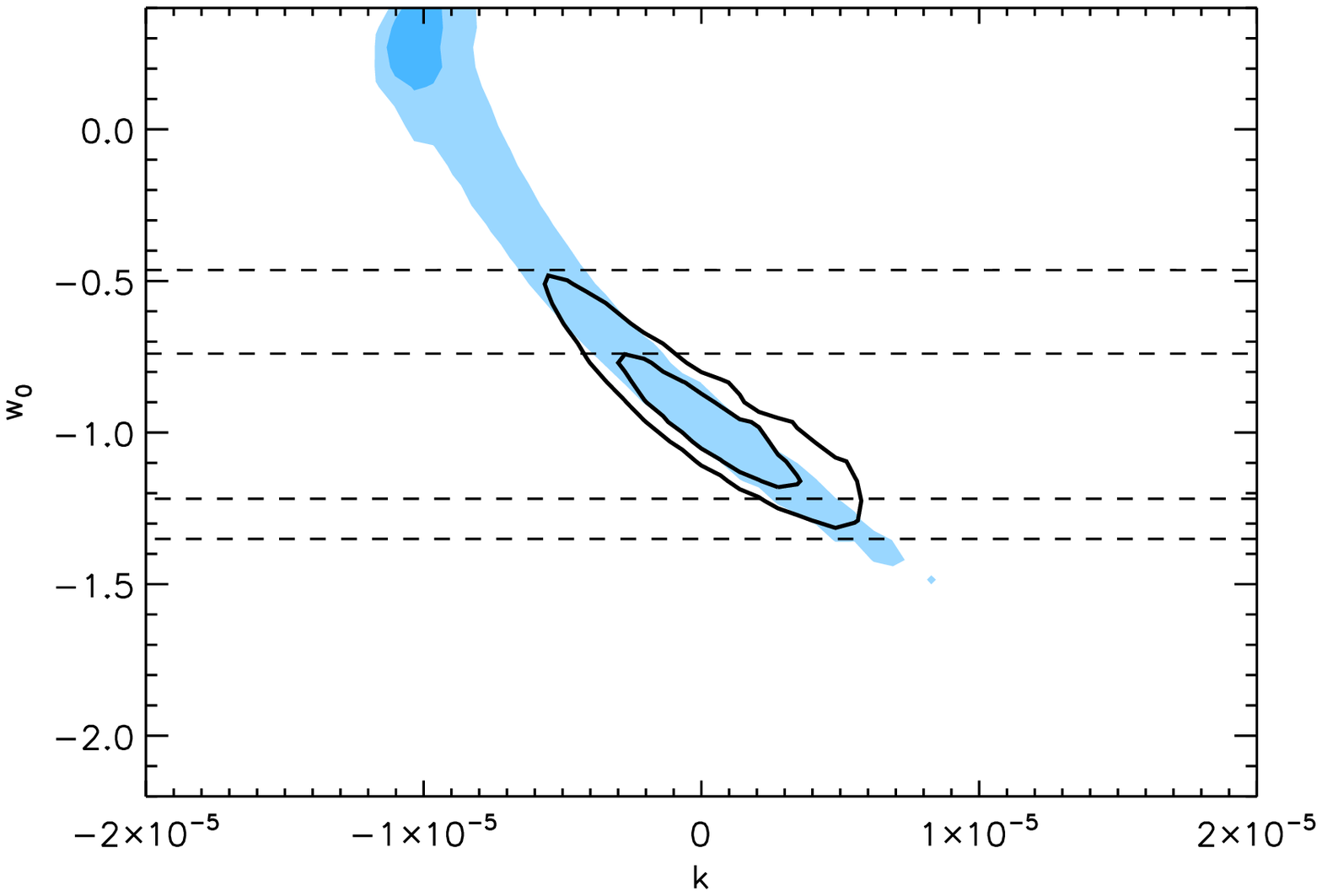} 
\includegraphics[width=8cm]{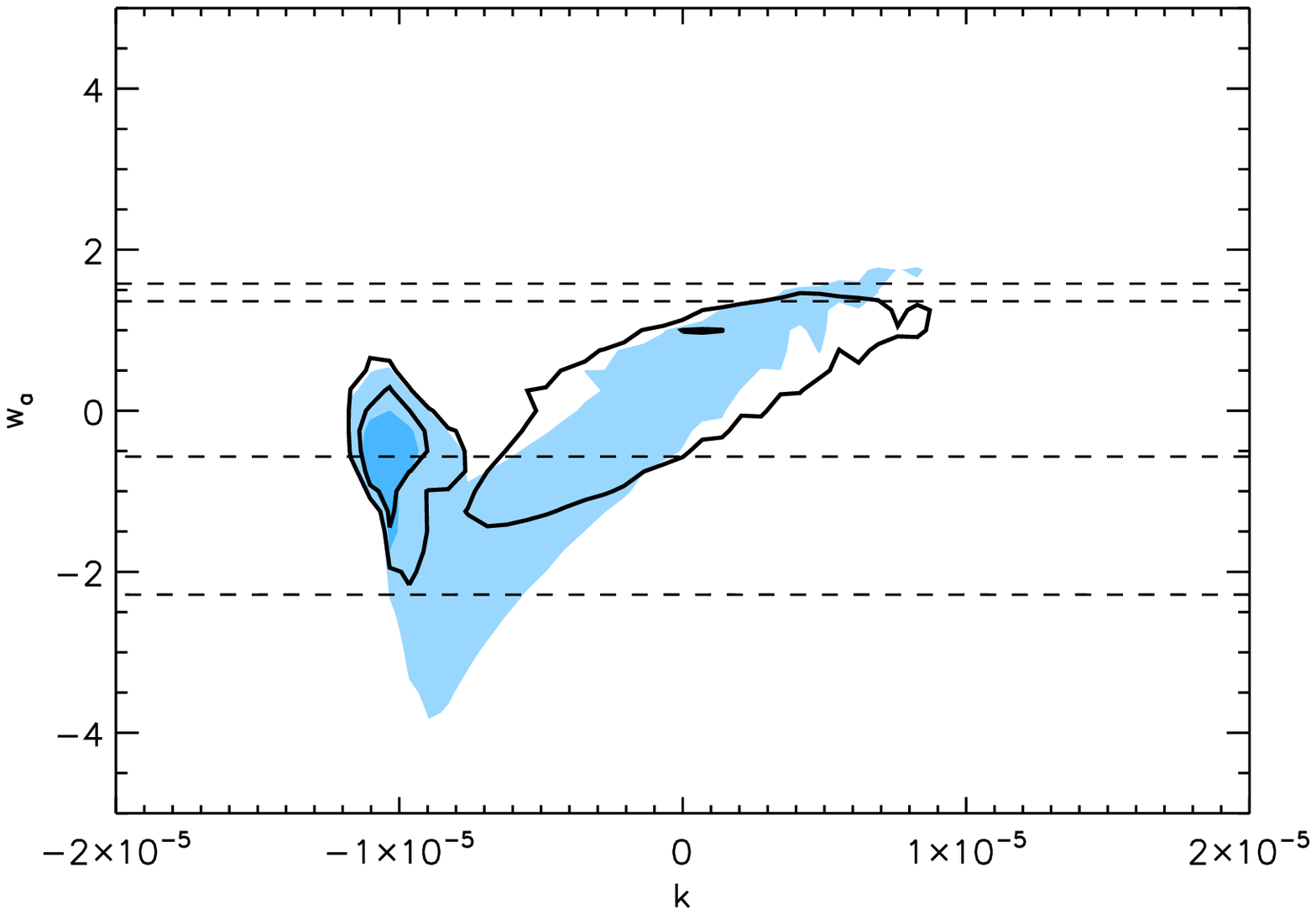}
\includegraphics[width=8cm]{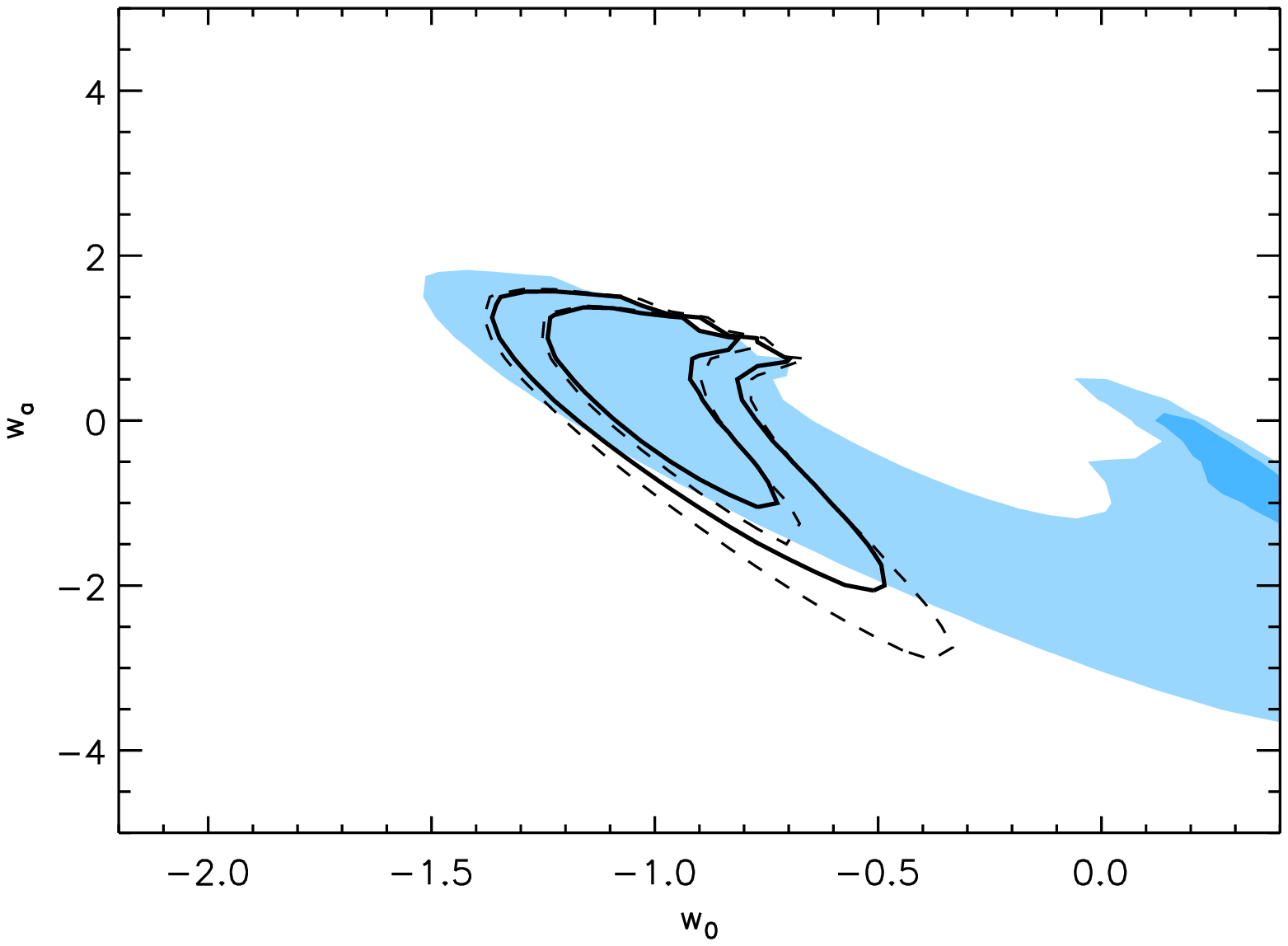} 
\includegraphics[width=8cm]{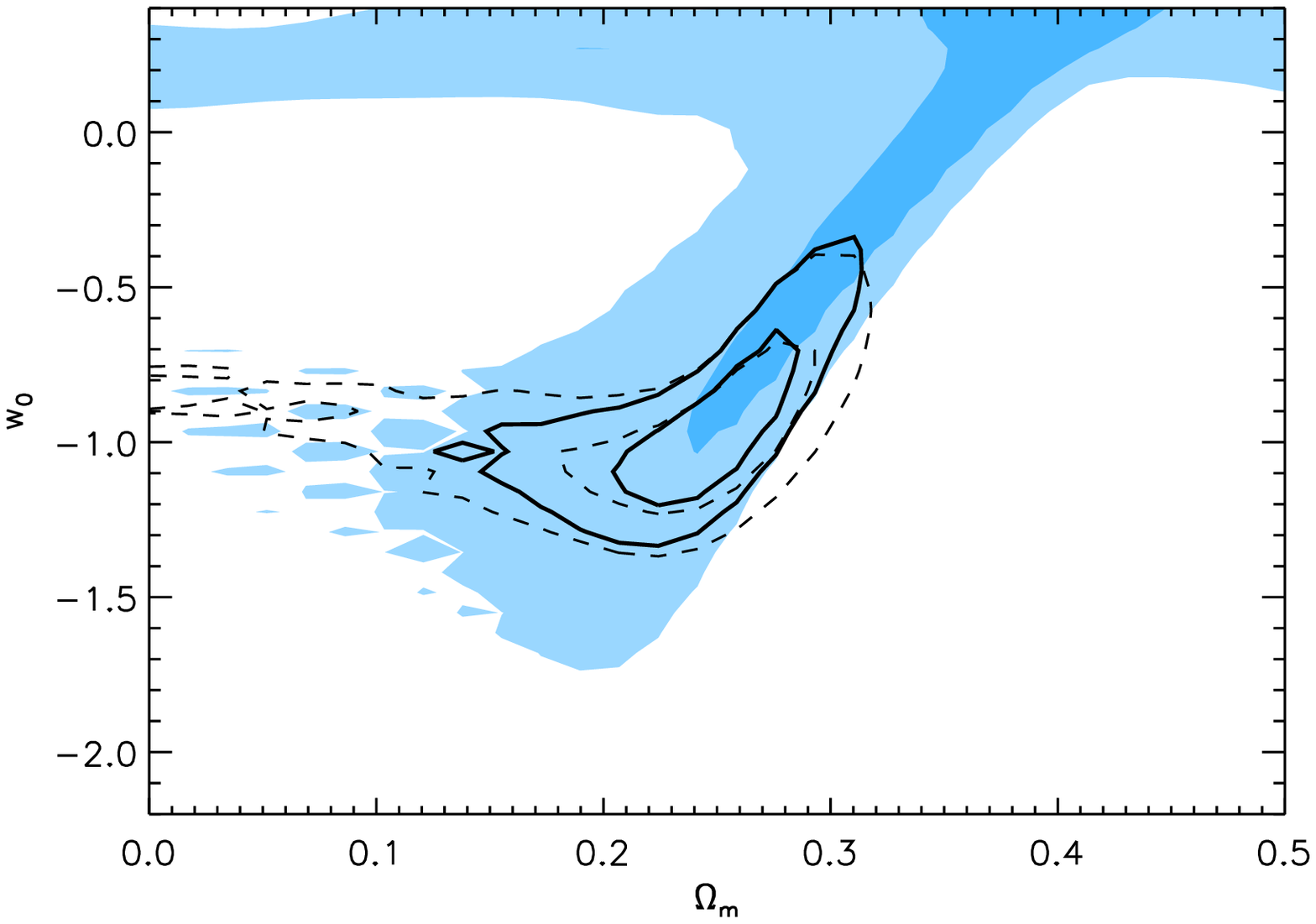} 
\includegraphics[width=8cm]{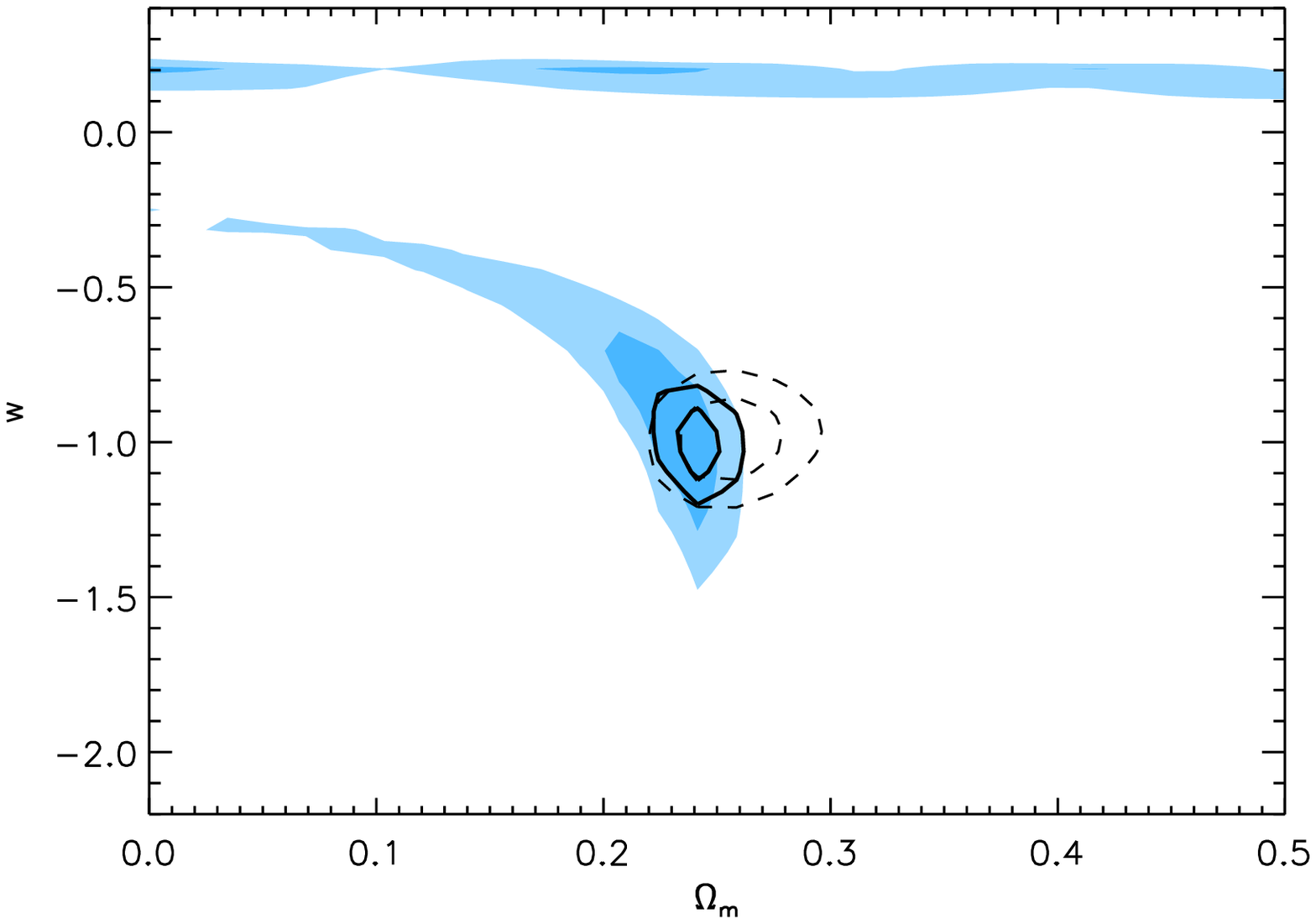} 
\includegraphics[width=8cm]{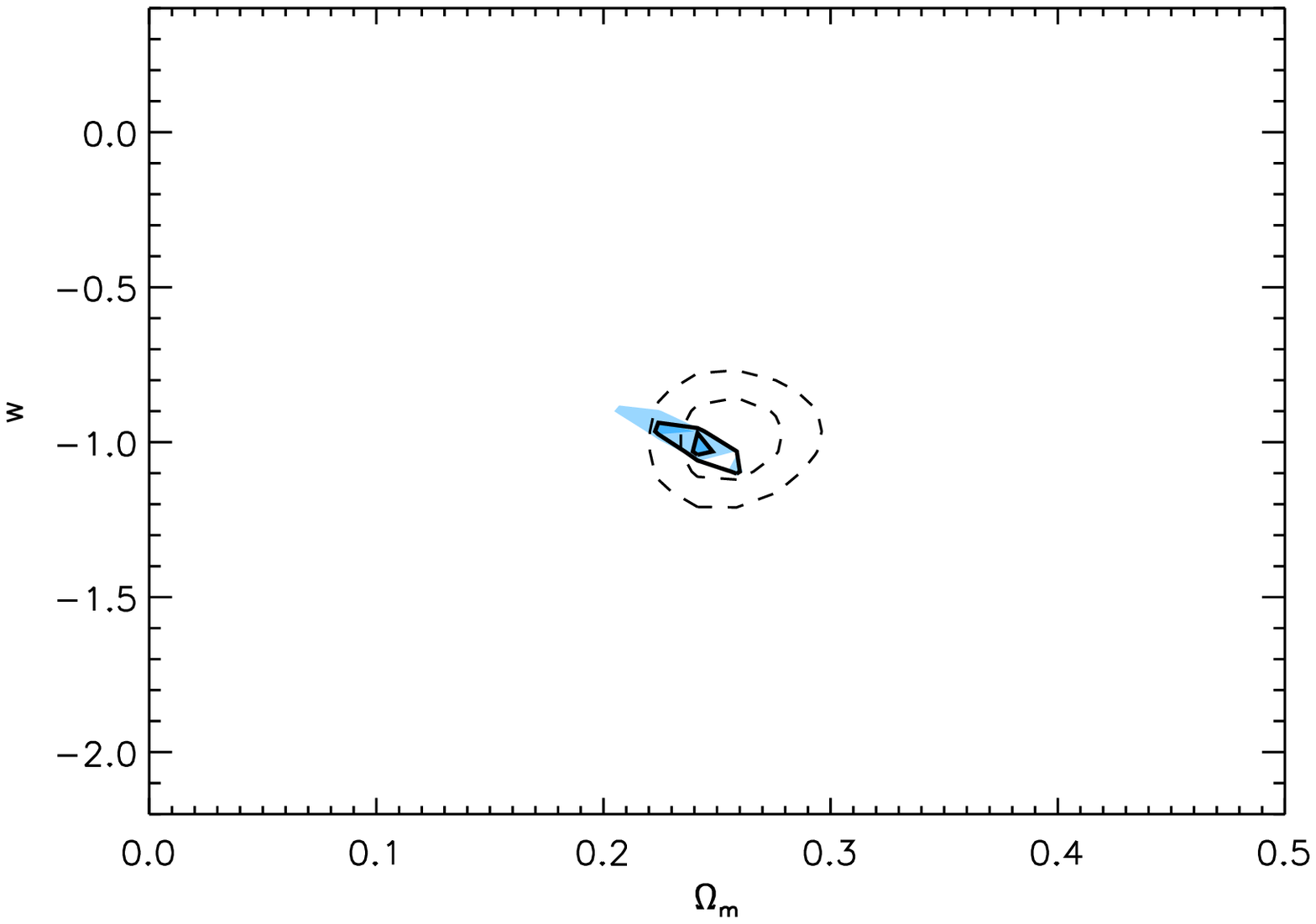}
\caption{As above but for TMT combined with a SNAP-like SN mission and Euclid 
BAO. Again, the additional $\sim 250$ SN in the range $1<z<3$ would lead to a 
significant improvement of the SN and joint constraints.}\label{TMTSNAPBAO}
\end{figure*}

Supernova measurements with Euclid (or SNAP), can only reach a maximum redshift of around $z\sim1.7$. However, the next generation of ground and space-based optical-IR telescopes will significantly extend this redshift range, which will bring about 
significant improvements in terms of dark energy constraints.  Specifically, JWST (through NIRcam imaging), should find about 50 supernovae and measure their light curves \cite{Riess}, and with E-ELT spectroscopy provided by HARMONI \cite{Thatte} the redshift and supernova type can be confirmed. The redshift range of this high-z sample is expected to be $1<z<5$. The redshift distribution of these supernovae is not easy to extrapolate, since even the most detailed current studies such as those of the SNLS team \cite{Perrett} only reach out to $z\sim1$. In the absence of a specific redshift distribution, we will simply assume it to be uniform in the above range. With these assumptions, our forecasts are shown in Fig. \ref{EELTSNAPBAO}. For comparison, we also  consider the alternative case of the TMT (also with JWST support) \cite{TMT}, which expects to find about 250 supernovae in the range $1<z<3$; this is shown in Fig. \ref{TMTSNAPBAO}. We empahsise that the numbers we use for the E-ELT and the TMT come from assumptions made in Phase A studies of their relevant instruments; the amounts of telescope time required for gathering each dataset are not necessarily comparable.

We can see that these high-redshift supernovae lead to significantly improved constraints, compared to the previous Euclid$+$SNAP case. On the other hand, the constraints from the E-ELT and the TMT are comparable, indicating that the larger redshift lever arm partially compensates the smaller number of supernovae. Note, however, that these improvements on SN constraints from the inclusion of high-redshift supernovae, will only have a moderate effect on the joint BAO+SN result, unless $k$ is independently constrained as discussed above.
Table \ref{table1} provides a comparison of the uncertainties in the various model parameters obtainable in each case. Our joint analysis of future BAO+SN constraints for the various cases shows that $k$ will only be probed at the level of $\sim\!10^{-5}$ (at 95\% confidence), which is below what is currently achievable using T(z) measurements\footnote{Note that the likelihood function for $k$ has two local maxima, one at $k \simeq\!-10^{-5}$ and one at $k=0$, so we do not show 1-$\sigma$ errors for $k$ in Table \ref{table1}.}. This also demonstrates the need to obtain independent determinations of T(z), which will break the degeneracy with $k$, having an important effect on improving the joint constraints on dark energy parameters. We will study this in detail in a follow-up publication.

\begin{table*}
\begin{tabular}{|c|c|c|c|c|}
\hline
Dataset & $\delta w_0$  & $\delta w_a$ & $\delta\Omega_m$ & $\delta k$ (2-$\sigma$) \\ 
 \hline
 \hline
Current (weak) & $0.25$ & $1.3$  & $0.06$ & $10^{-5}$ \\
Current (strong) & $0.22$ & $0.65$  & $0.06$ & $10^{-5}$ \\
\hline
Euclid(BAO)+SNAP & $0.15(0.35)$ & $0.4(1.6)$  & $0.03$ & $1.1\!\times\!10^{-5}$ \\
\hline
Euclid only (BAO+SN) & $0.15(0.35)$ & $0.6(1.6)$  & $0.03$ & $ - $ \\
\hline
Euclid(BAO+SN)+SNAP & $0.14(0.35)$ & $0.4(1.5)$  & $0.025$ & $9\!\times\!10^{-6}$ \\
\hline
Euclid(BAO)+SNAP+E-ELT & $0.13(0.3)$ & $0.4(1.45)$  & $0.023$ & $8\!\times\!10^{-6}$ \\
\hline
Euclid(BAO)+SNAP+TMT & $0.13(0.25)$ & $0.4(1.3)$  & $0.024$ & $8\!\times\!10^{-6}$ \\
\hline\hline
\end{tabular}
\caption{\label{table1}1-$\sigma$ (2-$\sigma$) uncertainty in the relevant model parameters, marginalising over the others, for the various datasets discussed. The uncertainty quoted for the photon-number violation parameter, $\delta k$, is at 2-$\sigma$. These constraints on dark energy parameters will become much stronger if $k$ is independently constrained by an external dataset, e.g. from $T(z)$ determinations.}
\end{table*}

\section{Conclusions}

We have studied two typical classes of phenomenological scenarios involving scalar fields coupled to radiation, specifically considering their effects on the redshift evolution of the cosmic microwave background temperature.
In the first of these a BSBM-type field provides a time variation of the fine-structure constant $\alpha$, while in the other the dynamical scalar field is responsible for the recent acceleration of the universe.

Our analysis shows that the effects of the coupling of scalar fields to photons, which include effects on luminosity distances, dramatically weaken current and future constraints on cosmological parameters. In particular, our results strongly suggest that in order to fully exploit forthcoming SN data one must also independently constrain photon-number non-conservation arising from the possible coupling of SN photons to the dark energy scalar field. In this context, direct measurements of the background temperature at different redshifts (such as those provided by ALMA and HIRES) can be used in combination with distance measures to break parameter degeneracies and significantly improve constraints on physical processes in the early universe.

Nevertheless, our analysis demonstrates that Euclid can, even on its own, provide useful dark energy constraints while allowing for the possibility of photon number non-conservation. Naturally, stronger constraints can be obtained in combination with other probes. In this context its worth emphasising that the only Euclid probes we considered are BAO and the proposed SN survey. The Euclid mission includes further probes which can be used to tighten the constraints. In this sense our results are conservative (but an analysis of this more general case is left for future work).

We have also considered the role of the increased redshift lever arm provided by type Ia supernovae at high redshift ($z>2$), such as can be found by JWST and ground-based extremely large telescopes. Specifically, we have considered two different samples which are meant to be representative of the E-ELT and TMT, with the former going deeper into the matter era while the latter has five times more supernovae. The constrains from both datasets (in combination with lower-redshift measurements) are quite comparable: the E-ELT provides a better constraint on the matter density (for which the increased redshift lever arm is the dominant factor) while the TMT provides better constraints on the dark energy parameters $w_0$ and $w_a$ (since, at least in the models we considered, dark energy is negligible at the higher redshifts, the larger number of supernovae provides the dominant effect).

Finally, let us point out that HIRES exquisite precision and stability will give it two other abilities of note: it will be able to make the first measurements of the cosmological redshift drift and also to map out the behaviour of fundamental couplings from about $z\sim0.5$ to $z\sim4$ and possibly well beyond. As discussed in \cite{Pauline1,Pauline3,Amendola}, both of these will provide further constraints on dynamical dark energy as well as key consistency tests of many of these scenarios. Thus the combination of Euclid and the E-ELT offers us the prospect of a complete mapping of the dynamics of the dark sector of the universe all the way up to redshift $z\sim4$, and our work highlights the point that mapping the bright sector of the universe---through $T(z)$ measurements---also plays a role in this endeavour.


\begin{acknowledgments} 
This work was done in the context of the project PTDC/FIS/111725/2009 from FCT (Portugal) and the cooperation grant `Probing Fundamental Physics with Planck' (PHC-EGIDE/Programa PESSOA, grant FCT/1562/25/1/2012/S), with additional support from Grant No. PP-IJUP2011-212 (funded by U. Porto and Santander-Totta). AA was supported by the Marie Curie grant FP7-PEOPLE-2010-IEF-274326 and a University of Nottingham Research Fellowship. We acknowledge useful comments and suggestions from Isobel Hook and other members of the Euclid Cosmology Theory and Transients \& Supernovae Science Working Groups. We also acknowledge use of the Planck Mission Cosmological Parameters Products, which are publicly available from the Planck Legacy Archive (http://www.sciops.esa.int/Planck).
\end{acknowledgments}

\bibliographystyle{JHEPmodplain}
\bibliography{models}

\end{document}